\documentclass[lettersize,journal]{IEEEtran}
\usepackage{amsmath,amsfonts}
\usepackage{algorithmic}
\usepackage{array}
\usepackage[caption=false,font=normalsize,labelfont=sf,textfont=sf]{subfig}
\usepackage{textcomp}
\usepackage{stfloats}
\usepackage{url}
\usepackage{verbatim}
\usepackage{graphicx}

\usepackage{multirow}
\usepackage[dvipsnames]{xcolor}
\usepackage{color}
\usepackage{cite}

\def\BibTeX{{\rm B\kern-.05em{\sc i\kern-.025em b}\kern-.08em
    T\kern-.1667em\lower.7ex\hbox{E}\kern-.125emX}}
\usepackage{balance}
\begin{document}
\title{CBSeq: A Channel-level Behavior Sequence For Encrypted Malware Traffic Detection}

\author{Susu Cui, Cong Dong, Meng Shen,~\IEEEmembership{Member,~IEEE,} Yuling Liu, Bo Jiang, and Zhigang Lu
\thanks{Manuscript received September 2, 2022; revised May 15, 2023. This work is supported by National Key Research and Development Program of China (No.2019QY1300), and the Strategic Priority Research Program of Chinese Academy of Sciences (No.XDC02040100). This work is partially supported by NSFC (No.61902376). This work is also supported by the Program of Key Laboratory of Network Assessment Technology, the Chinese Academy of Sciences, Program of Beijing Key Laboratory of Network Security and Protection Technology. \textit{(Corresponding author: Bo Jiang.)}}
\thanks{Susu Cui, Yuling Liu, Bo Jiang and Zhigang Lu are with the Institute of Information Engineering, Chinese Academy of Sciences, Beijing, China, and also with the School of Cyber Security, University of Chinese Academy of Sciences, Beijing, China (e-mail: cuisusu@iie.ac.cn, liuyuling@iie.ac.cn, jiangbo@iie.ac.cn, luzhigang@iie.ac.cn).}
\thanks{Cong Dong is with the Zhongguancun Lab, Beijing, China (e-mail: dongcong@zgclab.edu.cn).}
\thanks{Meng Shen is with the School of Cyberspace Security, Beijing Institute of Technology, Beijing, China (e-mail: shenmeng@bit.edu.cn).}
}

\markboth{Journal of \LaTeX\ Class Files,~Vol.~18, No.~9, September~2020}%
{How to Use the IEEEtran \LaTeX \ Templates}

\maketitle

\begin{abstract}
Machine learning and neural networks have become increasingly popular solutions for encrypted malware traffic detection. They mine and learn complex traffic patterns, enabling detection by fitting boundaries between malware traffic and benign traffic. Compared with signature-based methods, they have higher scalability and flexibility. However, affected by the frequent variants and updates of malware, current methods suffer from a high false positive rate and do not work well for unknown malware traffic detection. It remains a critical task to achieve effective malware traffic detection. In this paper, we introduce CBSeq to address the above problems. CBSeq is a method that constructs a stable traffic representation, behavior sequence, to characterize attacking intent and achieve malware traffic detection. We novelly propose the channels with similar behavior as the detection object and extract side-channel content to construct behavior sequence. Unlike benign activities, the behavior sequences of malware and its variant's traffic exhibit solid internal correlations. Moreover, we design the MSFormer, a powerful Transformer-based multi-sequence fusion classifier. It captures the internal similarity of behavior sequence, thereby distinguishing malware traffic from benign traffic. Our evaluations demonstrate that CBSeq performs effectively in various known malware traffic detection and exhibits superior performance in unknown malware traffic detection, outperforming state-of-the-art methods.
\end{abstract}


\section{Introduction}

\IEEEPARstart{C}{yber} attacks are increasingly frequent and severe. Attackers rely on malicious bots to perform credential stuffing attacks, denial-of-service attacks (DoS), and probe known vulnerabilities in web applications, generating large amounts of malware traffic~\cite{9519384}. To promptly detect malicious attacks and ensure network security, deploying a network intrusion detection system (NIDS) at the edge of the network is a key technology. The system passively monitors network traffic to detect malicious network activities. 

Early studies build signatures by extracting payloads from network traffic for malware traffic detection. However, traffic encryption technology has been widely used in various network services in recent years~\cite{DBLP:journals/tifs/ShenLZDH21,DBLP:journals/tifs/ShenWZW17}. To avoid the inspection of firewalls and NIDS, a large number of malicious attacks also use encryption technology to hide the content of the communication. Traditional signature-based detection does not work well since the payloads of encrypted traffic are random ciphertext. Moreover, it requires frequent updating of signature libraries and has great difficulty extracting signatures to cope with complex and unknown attacks.

In order to solve the problems of encrypted malware traffic detection, the current studies propose to use machine learning methods to identify malware traffic~\cite{9277523,9720753}. 
We summarize machine learning-based detection into two categories: (1) Statistics-based~\cite{DBLP:journals/sensors/VegaCHL20,9443025,Vu2022,DBLP:conf/ndss/MirskyDES18,DBLP:conf/kdd/AndersonM17,DBLP:conf/icnp/McGrewA16,DBLP:conf/esorics/CohenMKMEPS20}. It extracts the flow statistical features of network traffic, such as flow size and duration, and then it performs traffic detection by fitting the boundary between benign traffic and malware traffic. (2) Specific fields-based~\cite{roques2019detecting,DBLP:conf/iccns/DaiGLYLC19,10.1007/978-3-030-67090-0_8,DBLP:conf/ccs/TorroledoCB18,DBLP:conf/dsc/ZhaoLWHTC21,DBLP:journals/virology/AndersonPM18,DBLP:conf/ccs/AndersonM16}. It extracts specific plaintext in traffic and builds fingerprints or vulnerable features strongly related to malware traffic, such as TLS fingerprints, weak cipher suites, and destination port. And then, it performs traffic detection by determining whether the traffic matches malware fingerprints or vulnerable features.

Unfortunately, although machine learning-based methods have been extensively studied, they still struggle to achieve effective performance in real-world settings due to frequent malware variants and updates. Specifically, these methods rely on collected known malware traffic and extracted features for model fitting. However, in real-world environments, attackers usually repackage old malware and reuse known attack patterns, thus generating much unknown malware. Compared to old malware, updated unknown malware commonly employs various evasion methods, such as new transport protocols or tunnels, to escape NIDS inspection. Therefore, the unknown malware traffic changes the original statistical features (such as inter arrival time of packet, packet length), making it difficult for statistical-based methods to make effective judgments. In addition, attackers are also gradually avoiding fingerprint and vulnerable features. For example, attackers use cipher stunting to randomize SSL/TLS signatures to evade detection. As a result, this would make field-specific methods ineffective. Based on the above, there are still significant challenges in applying machine learning-based methods for encrypted malware traffic detection.

To address these challenges mentioned above, we investigate the network behavior of malware with the following aim: a stable traffic representation that can characterize unknown malware traffic generated by frequent variants and updates of malware. Our key finding is that, despite frequent updates and variants, the attacking intent\footnotemark[1] of malware often remains stable and displays solid inter-behavioral correlations. Taking Mirai~\cite{gopal2018mitigating} malware as an example, its attacking intent is to utilize infected IOT devices for launching DDoS attacks. During the execution of DDoS attacks, a significant amount of network traffic is generated, and these network connections are typically similar, such as having the same number of packets and similar inter arrival time between connections. Moreover, various Mirai variants, like Okiru and Satori, produce new attack payloads or modify protocols, cipher suites, and the addresses of their command \& control (C\&C) servers to evade traffic detection. However, the attacking intent of Mirai and its variants remains constant, that is, to utilize IOT devices for launching DDoS attacks. Additionally, the DDoS attack traffic of these variants also exhibits solid inter-behavioral correlations. Similarly, malware and its variants that execute online password cracking, worm propagation, and C\&C communication share the same characteristics: fixed attacking intent and solid inter-behavioral correlations.
\footnotetext[1]{Attacking intent refers to the purpose or goal behind a cyber attack carried out by malware. This can encompass a wide variety of objectives, such as data theft, denial-of-service attacks, causing damage to a system or network, gaining unauthorized access, or facilitating further attacks. In this study, we focus on the attacking intent that communicates with the Internet and generates significant network traffic.}


In this paper, we introduce the channel-level behavior sequence (CBSeq), a novel approach that constructs behavior sequence to achieve malware traffic detection, especially unknown malware traffic detection. Our core idea is aggregating traffic at the channel level and considering the channels with similar behavior as the detection object. First, unlike traditional methods, channel is the aggregation of multi-flows with the same source IP and destination IP into a whole, while traditional methods usually use flow as the granularity. Using channels allows us to explore rich behavior features more comprehensively. Second, we aggregate channels with similar activities and extract behavior sequences to characterize attacking intent. Despite the presence of a large amount of unknown malware, its attacking intent is relatively fixed. They typically generate similar network connections in a short period~\cite{9519384,garcia2015modelling}. Therefore, we construct behavior sequence from the perspective of attacking intent to improve the stability and unknown detection capability. CBSeq is performed in two steps.

\textbf{Constructing behavior sequence.} Behavior sequence is a novel traffic representation to characterize the attacking intent of malware. It is extracted from the side-channel content~\cite{arp2015torben} based on channels with similar activities to represent the solid inter-behavioral correlations, which are inherent to various malware and its updates and variants. We first extract abstract features from the channel, including duration, flow count, total data size, uplink data size, and downlink data size, to provide an overview of network activities. Then, clustering is performed based on channel abstract features to converge similar channels to the same cluster. The same cluster implies that the channels within the cluster have similar behavior. Next, we extract the channel sequence, which includes four kinds of sequences: packet number (PN) sequence, inter arrival time (IAT) sequence, source port (SP) sequence, and destination port (DP) sequence. The channel sequences of the same cluster are fused into the behavior sequence of the cluster to characterize the attacking intent. Finally, we transform the behavior sequence to behavior sequence embedding by word2vec, enhancing the representation of the sequence.

\textbf{Building MSFormer detector.} It is a Transformer-based malware traffic detection model. MSFormer builds independent sub-networks with the same structure for four types of sequences. It learns the internal relationships of elements within behavior sequence embeddings to distinguish malware traffic from benign traffic. Moreover, MSFormer does not require hand-crafted features and can achieve effective and accurate detection.

CBSeq is cross-protocol, robust, and capable of discovering unknown malware traffic. Specifically, CBSeq only uses the side-channel content of the traffic to construct behavior sequence. It does not involve application protocol information such as TLS features or HTTP features. Therefore, it enables encrypted and cross-protocol malware traffic detection. Meanwhile, we focus on channel-level analysis rather than traditional flow analysis, which has more robust feature representations. In addition, despite the variety of malware variants and updates, the attacking intent of similar malware tends to be stable. We construct behavior sequence based on attacking intent, which can detect unknown malware traffic generated by updates and variants.

\noindent{In summary, this paper provides the following contributions:}

\begin{itemize}
\item{We novelly consider channels with similar behavior as the detection object and construct behavior sequence to represent attacking intent.}
\item{We propose using word2vec to convert the behavior sequence into a meaningful numeric vector, which can enhance the inter-correlation of malware traffic.}
\item{We design MSFormer, a powerful Transformer-based multi-sequence fusion classifier. MSFormer builds independent sub-networks for different sequences. The sub-networks capture the sequence relationships based on the attention mechanism for effective and accurate detection.}
\item{We validate the effectiveness and efficiency of CBSeq through both known and unknown malware traffic detection scenarios. Compared with the baseline methods, CBSeq improves the AUC value by 1.4\% for known malware traffic detection and 16.0\% for unknown malware traffic detection. CBSeq has the highest performance compared to state-of-the-art methods.}
\end{itemize}

The remainder of this paper is structured as follows. Section II reviews related works and their limitations. Section III describes the preliminaries. Section IV introduces the design of CBSeq. Section V provides the experiment and evaluation. Section VI provides the conclusion.

\section{Related Work}
In this section, we review related works on malware traffic detection. We categorize these studies into three types: signature-based detection, statistics-based detection, and field-specific detection

\subsection{Signature-based Detection}
Signature-based detection methods extract signatures and rules that characterize attacks from malware traffic. Then, it achieves detection by matching them, such as string matching or regular expressions matching. Snort~\cite{DBLP:journals/ijics/GuptaS20} detects attacking or probing traffic through protocol analysis and content matching based on a predefined rule set. Snort can detect buffer overflow, port scanning, DoS, and many other attacks. Similarly, Suricata~\cite{DBLP:conf/sca2/ChibaAMOR19} checks network traffic with an extensive library of rules and signatures. It supports Snort rules and Lua scripts. And it has multi-threading, protocol detection, IP matching, file matching, and logging capabilities. In addition to the detection tools for industrial products, ~\cite{DBLP:journals/tifs/DongLCLC21} proposes a method to build signatures based on packet length. It constructs a behavior tree based on a sequence of packet length. And then, the behavior tree is the malicious signature to detect the encryption remote access Trojan.

On the whole, signature-based methods rely on predefined signature libraries for matching to achieve malware traffic detection. They are characterized by low false positive rates~\cite{DBLP:journals/asc/MasdariK20,DBLP:journals/tifs/DongLCLC21}. However, with the widespread use of encryption protocols, the extraction of signatures is more challenging due to the random ciphertext~\cite{DBLP:journals/cybersec/KhraisatGVK19,DBLP:conf/icct/IsingizweWLWWL21,10.1007/978-981-32-9949-8_48}. Moreover, signature-based methods face the challenges of malware updates and variants. Any modifications to existing malware, inclusive of new modules or changes to protocols, can render the original signatures obsolete. Consequently, these signatures require frequent manual updates, a task which is logistically demanding~\cite{DBLP:conf/acling/ApplebaumG021,DBLP:journals/jnsm/OtoumN21,10.1007/978-3-030-52856-0_2}. Furthermore, the efficacy of these signature libraries is inherently restricted to known malware traffic, rendering them impotent in the face of unknown malware traffic.

\subsection{Statistics-based Detection}
Statistics-based detection extracts flow statistical features and performs traffic detection by fitting the boundary between malware traffic and benign traffic. Recently, the studies~\cite{DBLP:journals/sensors/VegaCHL20,9443025,Vu2022,DBLP:conf/ndss/MirskyDES18} propose a large number of statistical features based on flow for distinguishing malware traffic and benign traffic, such as the packet number, flow size, inter arrival time, and duration. Beyond regular statistical features,~\cite{DBLP:conf/kdd/AndersonM17,DBLP:conf/icnp/McGrewA16} propose that the byte distribution of flow, packet size sequence, and packet time sequence can effectively detect malware traffic.~\cite{bader2022maldist} uses CNN and LSTM to learn the spatial and temporal features of raw flow bytes and the statistical information of the first 32 packets for malware traffic detection. To improve data imbalance and data drift, ~\cite{niu2022novel} employs an adaptive random forest (ARF) to learn the distribution features, plaintext information, and statistical features of the flow. In addition to conventional flow detection or packet detection,~\cite{DBLP:conf/esorics/CohenMKMEPS20} extracts the destination port sequence at the host level to characterize malicious activities, which can effectively detect malicious activities such as port scanning and worms. ~\cite{fang2021communication} proposes using channel traffic (packets composed of the same destination IP and destination port) as traffic granularity, extracting the distribution features, TLS handshake plaintext information, and statistical features for traffic representation, and using a random forest (RF) to enhance the malware traffic detection performance.

Many studies on statistics-based detection show extremely high performance. While being quite effective within a controlled dataset, statistics-based detection exhibits limited robustness when transposed to a real-world environment. This environment includes known traffic and much unknown malware traffic. The unknown malware traffic introduces new data distributions, which existing statistical methods struggle to accommodate, thereby compromising their accuracy~\cite{DBLP:journals/jsac/HanWZCYLSY21,DBLP:conf/kdd/AndersonM17,DBLP:conf/ccs/AndresiniPPLAC21,DBLP:conf/ndss/EdeBCRDLCSP20}. Although ~\cite{niu2022novel} uses an ARF to improve data drift, it requires the labels of new data to calculate the error of the model prediction results and update the model parameters accordingly. However, when it comes to detecting unknown malware traffic, the actual labels of the traffic cannot be known in advance, rendering this method unsuitable for unknown malware traffic detection. Furthermore, while ~\cite{fang2021communication} can gather more flows to improve the robustness of detection, it does not consider the behavior features of malware traffic and still conducts statistical analysis, which is unstable in the open-world environment. Hence, it cannot effectively detect unknown malware traffic.

\subsection{Field-specific Detection}
Field-specific detection extracts fingerprints or vulnerable fields strongly correlated with the attack behavior or attacker. And then, it trains a machine learning model to implement malware traffic detection. For example, JA3/JA3S extracted from TLS can be used as fingerprints for malicious tool identification~\cite{roques2019detecting}, while vulnerable features such as self-signed certificates and expired certificates can be used to detect anomalous traffic~\cite{DBLP:conf/iccns/DaiGLYLC19,10.1007/978-3-030-67090-0_8,DBLP:conf/ccs/TorroledoCB18,DBLP:conf/dsc/ZhaoLWHTC21}.~\cite{DBLP:journals/virology/AndersonPM18} introduces cipher suites, TLS extensions, and public key lengths to classify malware families.~\cite{DBLP:conf/ccs/AndersonM16} considers the contextual information of TLS flows, DNS responses, and HTTP headers and performs malicious family classification by them.

However, in the TLS v1.3 protocol~\cite{DBLP:conf/eurosp/KrawczykW16}, all handshake messages after ServerHello are encrypted. Therefore, they cannot check TLS v1.3 traffic and other protocol traffic. In addition, such methods face challenges due to the ability of update or variant malware to modify their TLS parameters, effectively evading detection. Thus, these methods demonstrate limited effectiveness against unknown malware traffic.

\section{Preliminaries}

In this section, we first introduce the traffic granularity and the definition of malware traffic detection. Then, we briefly introduce the self-attention mechanism.

\subsection{Traffic Granularity}
Network traffic is considered a collection of continuous packets. In order to analyze and detect the traffic, we need a discrete representation of the traffic, that is, to identify the granularity of traffic. In the existing traffic identification and detection works, the common traffic granularity mainly includes \textit{packet}, \textit{flow}, and \textit{channel}. According to~\cite{DBLP:journals/network/DainottiPC12,DBLP:journals/tifs/ShenZZXD21,DBLP:conf/ndss/MirskyDES18}, they are defined as:

\begin{itemize}

\item{\textit{Packet}: A single packet $p^{i}$ in continuous traffic is an identification object, which is defined as}
\begin{equation}
\label{equ1}
\begin{split}
p^{i} = (x^{i}, t^{i}, c^{i})\\
\end{split}
\end{equation}
where $x^{i}$ is the five-tuple information (source IP, source port, destination IP, destination port, transport layer protocol) of the $p^{i}$, $t^{i}$ is the time the packet is sent of the $p^{i}$ and $c^{i}$ is the payload of the $p^{i}$.

\item{\textit{Flow}: A flow $s^{i}$ is the traffic generated using the same five-tuple, or a bidirectional flow in which the source IP and destination IP are interchangeable. It is defined as:}
\begin{equation}
\label{equ2}
\begin{split}
s^{i} &= (x^{i}, t^{i}, P^{i})\\
P^{i} &= \{(x^{1}, t^{1}, c^{1}),\dots,(x^{m}, t^{m}, c^{m})\}
\end{split}
\end{equation}
where $x^{i}$ denotes the five-tuple information of $s^{i}$, $t^{i}$ denotes the start time of $s^{i}$, $P^{i}$ denotes the set of packets in $s^{i}$, $x^{1}=\dots=x^{m}$, $t^{1}<\dots<t^{m}$, and $m$ denotes the number of packets contained in $s^{i}$.

\item{\textit{Channel}: A channel $cn^{i}$ is the traffic generated by an IP pair, which is defined as:}
\begin{equation}
\label{equ3}
\begin{split}
cn^{i} &= (t^{i}, S^{i})\\
S^{i} &= \{(x^{1}, t^{1}, P^{1}),\dots,(x^{n}, t^{n}, P^{n})\}
\end{split}
\end{equation}
where $t^{i}$ denotes the start time of $cn^{i}$, $S^{i}$ denotes the flow set in $cn^{i}$. The source IP and destination IP in $S^{i}$ are the same or interchangeable. In $S^{i}$, $t^{1}<\dots<t^{n}$ , and $n$ denotes the number of flow contained in $cn^{i}$.
\end{itemize}

\subsection{Problem Definition}
In this paper, malware traffic refers to the network traffic that is generated by malware, and encrypted malware traffic detection monitors whether the channel traffic is benign or malicious. Channel is defined by Equation \eqref{equ3} mentioned above. In brief, we implement malware traffic detection in two phases. (1) Behavior sequence construction: it is a crucial process that aims to transform raw network traffic into a representation that can effectively characterize the network intent, particularly the attacking intent of malware. First, we extract abstract channel features and identify channels with similar network activities through clustering. Next, channel sequence features are extracted and aggregated for channels in the same cluster to build behavior sequences. Then, they are embedded using word2vec to transform the behavior sequences into vectors. This not only condenses the information contained in the sequences but also deeply characterizes the attacking intent of malicious activities. (2) Traffic detection: It distinguishes whether a channel cluster is benign traffic or malware traffic. The detection result for a channel cluster represents the detection results for all the channel traffic within that cluster. Note that we are not classifying the type of malware traffic but warning of any potential malware traffic.

Our goal is to design an end-to-end detection system to implement malware traffic detection and enhance the efficacy of NIDS. We assume that the NIDS is deployed on an edge device, which can replicate and forward incoming and outgoing network traffic packets to our detection system. Thus, the detection system only predicts traffic without intercepting any traffic, and it can be integrated into existing traffic defense systems as a plug-in. In addition, our system is suitable for detecting active attacks that attempt to access, compromise victim devices or retrieve sensitive data, such as brute force attacks, DoS attacks, and port scanning. However, it is not applicable to attacks that either do not generate traffic or only generate traffic within a local area network (LAN), such as network sniffing and privilege escalation.

\subsection{Self-Attention Mechanism}

\begin{figure}[!tbp]
    \centering
    \includegraphics[width=0.8\linewidth]{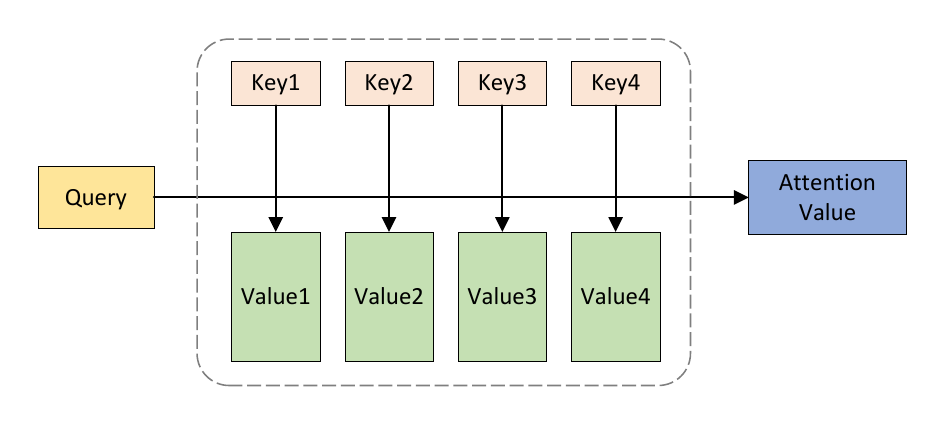}
    \caption{The attention mechanism. Given a \textit{query} vector (\textit{query} is an external query) related to the task. Attention value is obtained by computing the attention distribution of \textit{query} with \textit{key} and attaching it to the \textit{value}.}
    \label{attention}
\end{figure}

The universal approximation theorem states that both feedforward and recurrent neural networks have strong capabilities. However, the model becomes more complex when the neural network has too much input information. Moreover, the problem of long-distance dependence in recurrent neural networks persists. The attention mechanism~\cite{DBLP:conf/nips/MnihHGK14} is introduced to improve the efficiency of neural networks. It refers to the attention mechanism of the human brain and selects only some key information inputs for processing.

The essence of the attention mechanism is an addressing process, as shown in Fig.~\ref{attention}. The process begins with a \textit{query} vector that corresponds to a task-specific request. The \textit{query} is compared with each \textit{key}, a representation of the different parts of the input data, to generate an attention distribution. This distribution dictates the amount of attention allocated to each part of the input, based on its relevance to the \textit{query}.

Subsequently, this attention distribution is used to weigh the associated \textit{value}, another form of input data representation. This ensures that the most relevant parts of the input to the \textit{query} significantly influence the final outcome. The weighted \textit{values} are aggregated to generate the attention mechanism's output, a refined representation of the input data based on the \textit{query}. Thus, the attention mechanism effectively tailors the focus on input data for each unique \textit{query}, enhancing its overall performance.

The self-attention mechanism~\cite{DBLP:conf/nips/VaswaniSPUJGKP17}, a variant of the attention mechanism, reduces the dependence on external information and better captures the internal relevance of features. In self-attention, each input vector has three different vectors, which are \textit{query} vector ($\boldsymbol{Q}$), \textit{key} vector ($\boldsymbol{K}$), and \textit{value} vector ($\boldsymbol{V}$). First, $\boldsymbol{Q}=[\boldsymbol{q}_{1},\dots,\boldsymbol{q}_{N}]$, $\boldsymbol{K}=[\boldsymbol{k}_{1},\dots,\boldsymbol{k}_{N}]$, $\boldsymbol{V}=[\boldsymbol{v}_{1},\dots,\boldsymbol{v}_{N}]$ are obtained from the input sequence $\boldsymbol{X}=[\boldsymbol{x}_{1},\dots,\boldsymbol{x}_{N}]$. $\boldsymbol{q}_{i}$ , $\boldsymbol{k}_{i}$, $\boldsymbol{v}_{i}$ are defined as:

\begin{equation}
\label{equ4}
\begin{split}
\boldsymbol{q}_{i} &= \boldsymbol{W}_{q}\boldsymbol{x}_{i}\\
\boldsymbol{k}_{i} &= \boldsymbol{W}_{k}\boldsymbol{x}_{i}\\
\boldsymbol{v}_{i} &= \boldsymbol{W}_{v}\boldsymbol{x}_{i}
\end{split}
\end{equation}
where $\boldsymbol{W}_{q}$, $\boldsymbol{W}_{k}$, $\boldsymbol{W}_{v}$ are the parameter matrixs of the linear mapping.

We calculate the dot product between $\boldsymbol{Q}$ and $\boldsymbol{K}$. Then, to prevent the result from being too large, we divide it by a scale $\sqrt[]{d_{k}}$, where $\sqrt[]{d_{k}}$ is the dimension of $\boldsymbol{q}_{i}$ and $\boldsymbol{k}_{i}$. Next, we use Softmax to normalize the result to a probability distribution. Finally, we multiply result by the matrix $\boldsymbol{V}$ to obtain a weight summation representation. The operation is defined as:
\begin{equation}
\label{equ5}
\begin{split}
Attention(\boldsymbol{Q},\boldsymbol{K},\boldsymbol{V})=softmax(\frac{\boldsymbol{Q}\boldsymbol{K}^{T}}{\sqrt[]{d_{k}}})\boldsymbol{V}
\end{split}
\end{equation}

\begin{figure*}[!tbp]
    \centering
    \includegraphics[width=1\textwidth]{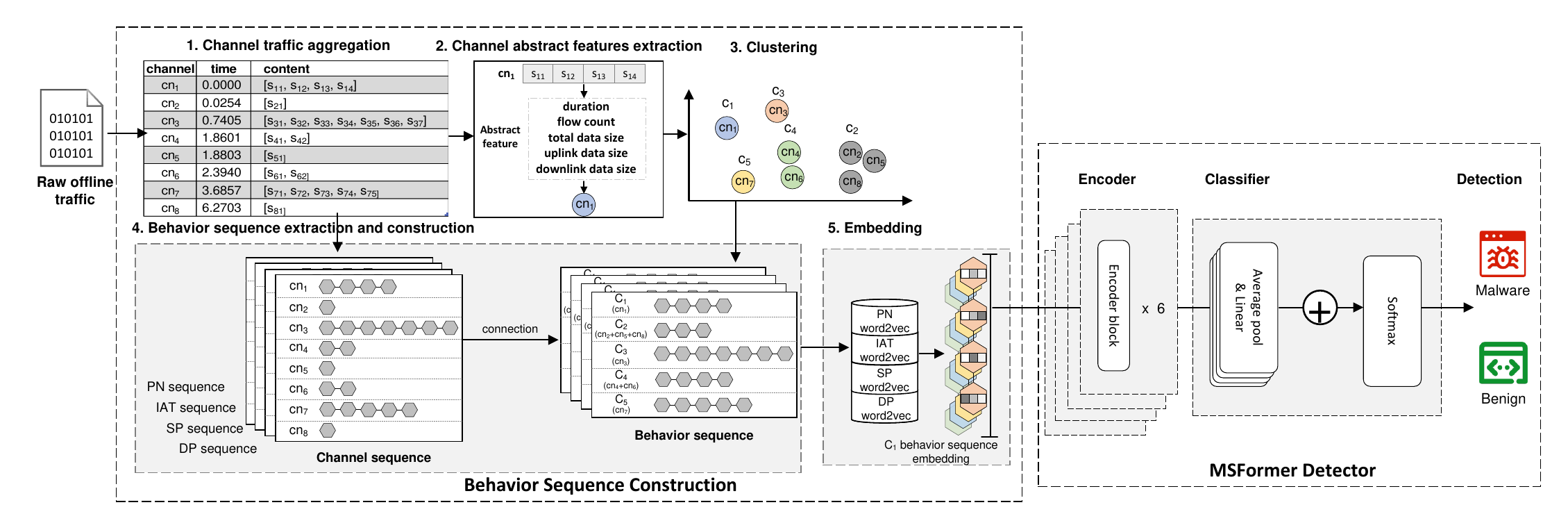}
    \caption{The framework of CBSeq. It includes \textit{behavior sequence construction} and \textit{MSFormer detector}. First, \textit{1. channel traffic aggregation} is conducted on the raw offline traffic. Then, we execute \textit{2. channel abstract features extraction} and utilize the abstract features to implement \textit{3. clustering}, thereby clustering similar channel traffic into a singular cluster. In the \textit{4. behavior sequence extraction and construction}, we extract the four types of sequence to compose the channel sequence. After that, the channel sequences from the same cluster are connected to formulate the behavior sequence. In \textit{5. Embedding}, we enhance the representation of the behavior sequence to generate the behavior sequence embedding, which is the input for the MSFormer Detector. Finally, the MSFormer detector distinguishes malware traffic from benign traffic.}
    \label{framework}
\end{figure*}

\section{Method}

In this section, we present the design of CBSeq. Firstly, we introduce the core idea. Secondly, we describe the process of behavior sequence construction and give the proposed MSFormer detector.

\subsection{Core Idea}

In this paper, we introduce CBSeq, and the framework is shown in Fig.~\ref{framework}. It first characterizes the attacking intent of malware traffic by behavior sequence construction. And then, it builds a detector named MSFormer to distinguish malware traffic from benign traffic. Our key finding is that, despite malware updates and variants having different functional modules, their attacking intent tends to remain fairly fixed. They typically generate a large number of similar network activities in a short period, such as targeting the same victim host with a large number of connections for online password cracking and DoS attacks and sending the same worm-containing emails to a large number of email addresses. Overall, the activities between the attacking host and the victim host exhibit strong inter-behavioral correlations. 

Therefore, we propose a behavior sequence for characterizing attacking intent. Firstly, we perform traffic aggregation based on the channel, which aggregates the traffic between the attacking host and the specific victim host. Since the network activities of the malware are similar, it also indicates that the flows within the channel have similar behavior features among themselves. In addition, attacks like worms usually present similar network activities between the attacking host and multiple victim hosts. Thus, we group similar channels together based on channel abstract features, including duration, flow count, total data size, uplink data size, and downlink data size. Channels within the same cluster have similar behavior features among themselves. In order to characterize the solid intre-behavioral correlations of the malware traffic, firstly, we extract four types of sequence of channel traffic: PN sequence, IAT sequence, SP sequence, and DP sequence, which are defined as follows.
\begin{itemize}
    \item \textit{PN sequence}: The sequence consists of the number of packets sent by the client in each flow within the channel.
    \item \textit{IAT sequence}: The sequence consists of the inter arrival time between flows in the channel. It is the subtracted value of the current flow's start time and the previous flow's start time. Thus, the first value of the time interval sequence is always 0.
    \item \textit{SP sequence}: The sequence consists of the source ports of each flow in the channel.
    \item \textit{DP sequence}: The sequence consists of the destination ports of each flow in the channel.
\end{itemize}

Secondly, we join and fuse the sequences of channels in the same cluster and obtain the behavior sequence. Unlike benign traffic, behavior sequences of malware traffic have solid intre-behavioral correlations and similarities. Moreover, in order to deeply capture the behavior features and attacking intent inside the sequences, we introduce word2vec to convert the behavior sequence into behavior sequence embedding.

Based on the behavior sequence, we build an MSFormer detector to distinguish malware traffic from benign traffic. MSFormer is a Transformer-based multi-sequence fusion classifier. It builds independent sub-networks for four sequences to learn the behavior embeddings, and the sub-networks capture the sequence information based on the attention mechanism. Finally, it fuses the sequences together and builds a Softmax classifier to achieve effective and accurate malware traffic detection, especially unknown malware traffic detection.

In addition, CBSeq effectively counters evasion techniques employed by malware. First, CBSeq's focus on traffic behavior analysis to characterize the attacking intent provides more stability against malware evasion techniques than traditional features. Moreover, CBSeq uses side-channel content like packet count and inter arrival time, which are accessible under encryption and tunnel techniques. Finally, CBSeq does not rely on TLS handshake information, thereby effectively identifying malware traffic even when the malware uses random cipher suites.

\subsection{Behavior Sequence Construction}
In this section, we describe the process of behavior sequence construction. First, we perform channel traffic aggregation and extract channel abstract features to represent the channel's network activity overview. The channel abstract features are then used for clustering to discover channels with similar activities and we perform behavior sequence extraction and construction from the aggregated channel clusters. Finally, the behavior sequence is converted into behavior sequence embedding.

\subsubsection{Channel Traffic Aggregation}
Network traffic is a collection of continuous packets. To perform traffic analysis and detection, we first split the continuous traffic according to the time window of 24 hours. We set this 24-hour window considering the daily activity similarity and the time preference bias. First, some infected hosts, such as the host infected by Zeus malware, could maintain communication with a specific server for several days. Despite this extended duration, the behavior exhibited within every 24 hours tends to be similar. As a result, splitting the traffic into 24-hour windows did not lead to a loss of significant behavior patterns. Second, we consider the tendency of certain malware to exhibit activity more frequently during specific times of the day. For instance, some malware may be programmed to be more active during nighttime hours. Using a 24-hour window, we could capture these daily fluctuations in activity, thereby avoiding this time preference bias and providing a more comprehensive view of the malware's network behavior. 

Next, in order to analyze and detect the traffic, it is primary to determine the granularity of the traffic. We aggregate the daily traffic according to the channel. The traffic from different channels is arranged by the start time, and each channel usually contains multiple flows. 

\subsubsection{Channel Abstract Features Extraction}
We extract abstract features from the channel traffic. According to the malware's function, the network activities generated by the infected hosts are varied. However, despite the wide variety of malware and attack activities, they can be summarized and categorized in terms of traffic statistics. We summarize and focus on the following two types of attacks.

\begin{itemize}

\item{\textit{Single-node persistent attack.} It mainly targets a specific server for port scanning or access. For example, Trickbot sends messages to the 
C\&C server periodically~\cite{DBLP:conf/kes/GittinsS20}. This type of attack usually results in a large number of connections between the infected host and the specific server, and the connections behave similarly. At the traffic level, it manifests itself as channel traffic containing multiple flows inside and with similar behavior between flows. For example, there is a greater similarity in the inter arrival time, duration, amount of data transferred, and ports accessed by the flows.}

\item{\textit{Multi-node transient attack.} It mainly targets many victim hosts for malware propagation. Unlike the single-node persistent attack, in the multi-node transient attack, the channel traffic within the infected host consists of fewer flows, and there is rarely similar behavior between flows. However, there is a similar behavior between the flows from different channels. For example, Geodo works together with a worm that uses email as an attack vector, which obtains a large number of target email addresses from the C\&C server and sends similar emails to these email addresses to spread malware~\cite{kuraku2020emotet}. Therefore, at the traffic level, there is a large similarity in the amount of data transferred, flow count, and duration between different channels.}

\end{itemize}

Since the channel traffic generated by the single-node persistent attack has similar behavior among its internal flows, the attacking intent can be characterized by the single-channel traffic. However, the multi-node transient attack mainly manifests itself as similar behavior among multiple channels. Therefore, it is required to correlate similar channel traffic to characterize the attacking intent.

To mine the network behavior of the multi-node transient attack, we first extract the abstract features from the channel traffic. We use duration, flow count, total data size, uplink data size, and downlink data size as channel abstract features and consider them as the overview of the channel.

\begin{figure*}[!tbp]
    \centering
    \includegraphics[width=\textwidth]{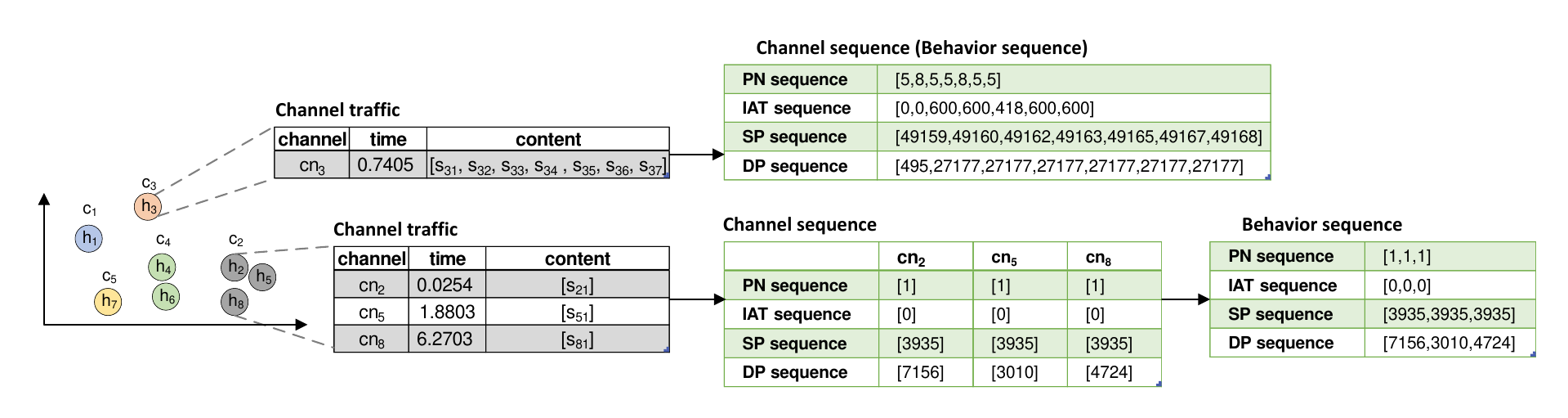}
    \caption{The example of behavior sequence extraction and construction from channel cluster. }
    \label{sequence}
\end{figure*}

\subsubsection{Clustering}
To discover channels with similar behavior, we cluster the channels based on the similarity of the abstract feature. First, we use the density-based spatial clustering of applications with noise (DBSCAN)~\cite{DBLP:conf/icadiwt/KhanRAFSV14} algorithm for channel clustering. Since attack activities are usually completed within a specific time interval, we consider similar activities of malicious channels to be time-limited. For this reason, we slice the clusters based on time windows such that individual channels of the same cluster belong to the same time window. 

DBSCAN is a basic algorithm for density-based clustering. It can discover clusters of different shapes and sizes from a large amount of data containing noise and outliers. The DBSCAN algorithm uses two parameters:

\begin{itemize}
    \item{MinPts: It refers to the minimum number of points (a threshold) needed to be clustered together for a region to be considered dense.}
    \item{Eps: It is a distance measure that is used to locate points in the neighborhood of any point.}
    
\end{itemize}

The purpose of clustering in this paper is to group together channels exhibiting similar behavior during multi-node transient attacks. If the eps is large, it will lead to both malware and benign traffic in the same cluster. Therefore, the eps setting should be small to ensure a single traffic label in the same cluster. Meanwhile, the single-node persistent attack mainly shows the high similarity within the channel and the low similarity between different channels. Therefore, this should be a separate cluster for each channel traffic. More details are shown in Section~\ref{sec:params}.

\subsubsection{Behavior Sequence Extraction and Construction}

After clustering, we extract the behavior sequences of channel clusters to characterize the network behavior and attacking intent, and an example of the process is shown in Fig.~\ref{sequence}. First, the clustering in the previous step divides the channels into different clusters. It is worth noting that the clustering divides all the channels into different clusters, including the clusters containing only a single channel. Then, for each cluster, we sort the channels by their start times. Next, we extract sequences for each channel, including PN sequence, IAT sequence, SP sequence, and DP sequence, and connect each channel sequence of the same cluster to form the behavior sequence.

For a cluster containing only one channel, the extracted channel sequence represents the cluster behavior sequence. And for the clusters containing multiple channels, the extracted channel sequences are further connected in the time order of channels to construct the behavior sequence. Regardless of whether the clusters consist of a single channel or multiple channels, it is evident from the example that the aforementioned four types of sequences bear internal similarities. Therefore, we consider that the above four types of sequences can effectively reveal the attacking intent.

\subsubsection{Embedding}

Although the packet number, inter arrival time, source port, and destination port are numerically represented, the raw numbers cannot better represent the channel's activity intent. The reasons are summarized as follows. (1) Packet numbers exhibit significant differences. The packet number takes values from 1 to 3203169 in the observed traffic. Although normalization can address the problem influenced by significant numerical differentiation, in theory, there is no maximum value for packet numbers. This could considerably impact the normalization results due to the data distribution of the actual traffic. Therefore, this will directly reduce the detection results of the model. (2) Despite the numerical representation, the source port and destination port do not have similar meanings between closer numbers~\cite{DBLP:conf/esorics/CohenMKMEPS20}. For example, port 80 and port 8080 indicate the same web service, though they are far apart. However, port 22 and port 23 are closely spaced, while port 22 is used for SSH and port 23 is used for Telnet. Therefore, to better capture the behavior features and activity intent within the sequence, we employ word embedding to represent the original behavior sequence.

Word embedding has been successfully used for various natural language processing tasks and speech processing~\cite{DBLP:journals/corr/abs-1901-09069}. It maps each word into a dense vector in a low-dimensional space (50-300 dimensions). Semantically similar words are more similar in the vector space. Word2vec is a widely used algorithm in word embedding. The core idea of word2vec is predicting the relationship between each word and its contextual words. Word2vec has two algorithms, a skip-grams algorithm and a continuous bag of words (CBOW) algorithm~\cite{DBLP:journals/corr/abs-1301-3781}. In this paper, we use the CBOW algorithm to train the word2vec model, and CBOW aims to predict a word given its context.

In this paper, we transform four types of sequences into embedding representations through the trained word2vec models. Each number in a sequence is considered a word, and the whole sequence is considered a sentence. We train the corresponding word2vec models for the PN sequences, IAT sequences, SP sequences, and DP sequences, respectively. Each number within the sequences is converted into a meaningful numerical vector. Meanwhile, within the same type of sequence, the output vectors have higher similarity than the original input numbers, which addresses the issue of significant differences in the numbers and also characterizes the intent of the malware's network activity.

\begin{table*}[!t]
\renewcommand\arraystretch{1.2}
\setlength\tabcolsep{4px}
\centering
\caption{The Statistics of experimental datasets.}
\begin{tabular}{ccccccccc}
\hline
\multirow{2}*{Dataset}&\multirow{2}*{Class}& Channel Traffic  & Samples Number in  &Samples Number in& Unique Source &	Unique Destination \\
 &&Number&Known Detection &Unknown Detection &Port Frequency& Port Frequency\\
\hline
Benign-ALL&	Benign&	32860 & 3965 & 32860 &	1.68\%&	0.00\%\\

\multirow{6}*{CTU-6}&Zeus&	6548 & 6548&\multirow{6}*{16430} &\multirow{7}*{6.93\%}&\multirow{7}*{17.21\%}\\
&Emotet&2864&2864&&&\\
&Miuref&1286&1286&&&\\
&Trickbot&	6314&	6314&&&\\
&Dridex&	6320&	6320&&&\\
&Downloadguide&	457&	457&&&\\
CTU-ALL&Unknown Malware&75781&-&16430&&\\
\hline
\end{tabular}
\label{dataset}
\end{table*}

\subsection{Proposed MSFormer Detector}

MSFormer is a model based on Transformer~\cite{DBLP:conf/nips/VaswaniSPUJGKP17} for malware traffic detection. Its framework is shown in Fig.~\ref{framework}. The input of the MSFormer is behavior sequence embedding, and the encoder module of Transformer is used to deep mine each of the four types of sequences. It learns the relationship between elements inside the sequences and captures the internal behavior features of the sequences. Finally, MSFormer classifies the output sequence of the encoder based on Softmax to achieve malware traffic detection. MSFormer consists of input layer, encoder, and classifier.


\subsubsection{Input Layer}
The input of MSFormer is the behavior sequence embedding. In this paper, we utilize behavior sequence extracted based on channel-level to characterize attacking intent. The behavior sequence contains PN sequence, IAT sequence, SP sequence, and DP sequence. Next, we use the trained word2vec model to transform the original values of sequence elements into word embedding, thereby forming the behavior sequence embedding. Employing word2vec can more deeply reveal the channel's internal relationships.

\subsubsection{Encoder}
Encoder contains four sub-encoders with identical structures, each consisting of six same encoder blocks. Encoder performs deep learning on the behavior sequence embedding from the input layer. The encoder block consists of two sub-layers, the multi-head attention, and the feed-forward network. In addition, each sub-layer has an Add \& Norm module.

Multi-head attention is a module for attention mechanisms that runs through an attention mechanism several times in parallel. The independent attention outputs are then concatenated and linearly transformed into the expected dimension. Intuitively, multi-head attention allows for attending to parts of the sequence differently. It is defined as:

\begin{small}
\begin{equation}
\label{equ6}
\begin{split}
MultiHead(\boldsymbol{Q},\boldsymbol{K},\boldsymbol{V}) &= Concat(head_{1},\dots,head_{h})\boldsymbol{W^{o}}\\
head_{i} &= Attention(\boldsymbol{QW_i^q},\boldsymbol{KW_i^k},\boldsymbol{VW_i^v})
\end{split}
\end{equation}
\end{small}
where $\boldsymbol{W^{o}}$, $\boldsymbol{W_i^q}$,$\boldsymbol{W_i^k}$ and $\boldsymbol{W_i^v}$ are all learnable parameter matrices.

Next, the second sublayer is a fully connected feed-forward network, consisting of two linear transformations with rectified linear unit (ReLU) activation in between. It is defined as:

\begin{equation}
\label{equ7}
\begin{split}
FFN(x) = ReLU(W_{1}x+b_{1})W_{2}+b_{2}
\end{split}
\end{equation}
where $W_{1}$ and $W_{2}$ are the weight parameters, $b_{1}$ and $b_{2}$ are the bias parameters.

Add \& Norm is connected after each sublayer in each encoder block. Specifically, Add represents the residual connection. It is the same as the residual connections in other neural network models, which are used to transfer the information deeper and enhance the fitting ability of the model. Norm represents the normalization layer. As the network's layer count increases, the parameters may enlarge or exhibit greater variances after computations across multiple layers. This leads to anomalies in the learning process and very slow convergence of the model. Therefore, normalization can improve the performance of the model.

\subsubsection{Classifier}
The classifier converts the output sequence into detection probabilities. First, it performs an average pooling of the output sequence to produce a single vector that characterizes the whole sequence. Then, this vector is mapped to a two-dimensional vector using the linear layer. Finally, the four vectors generated for the four types of sequence are summed, and Softmax is used to generate detection probabilities for encrypted malware traffic detection.

\section{Experiment and Evaluation}

In this section, we conduct the experiment and evaluation. Firstly, we describe the experiment settings, including evaluation traffic, parameter tuning, setting of the CBSeq, and baseline methods. We focus on evaluating the performance of CBSeq for both known malware traffic detection and unknown malware traffic detection. In addition, we perform analysis on embedding and analysis on behavior sequence to comprehensively evaluate CBSeq.

\subsection{Experiment Settings}

\subsubsection{Evaluation Dataset}

To perform and evaluate CBSeq, we use a traffic capture tool to store the mirrored data from the switches at the enterprise network offline in a traffic database as PCAP files. In the traffic database, we randomly selected 28 local devices whose traffic data is generated by communicating with the internet from the benign traffic dataset. The dataset is named Benign-ALL.

\begin{figure}[!t]
    \centering
    \includegraphics[width=0.9\linewidth]{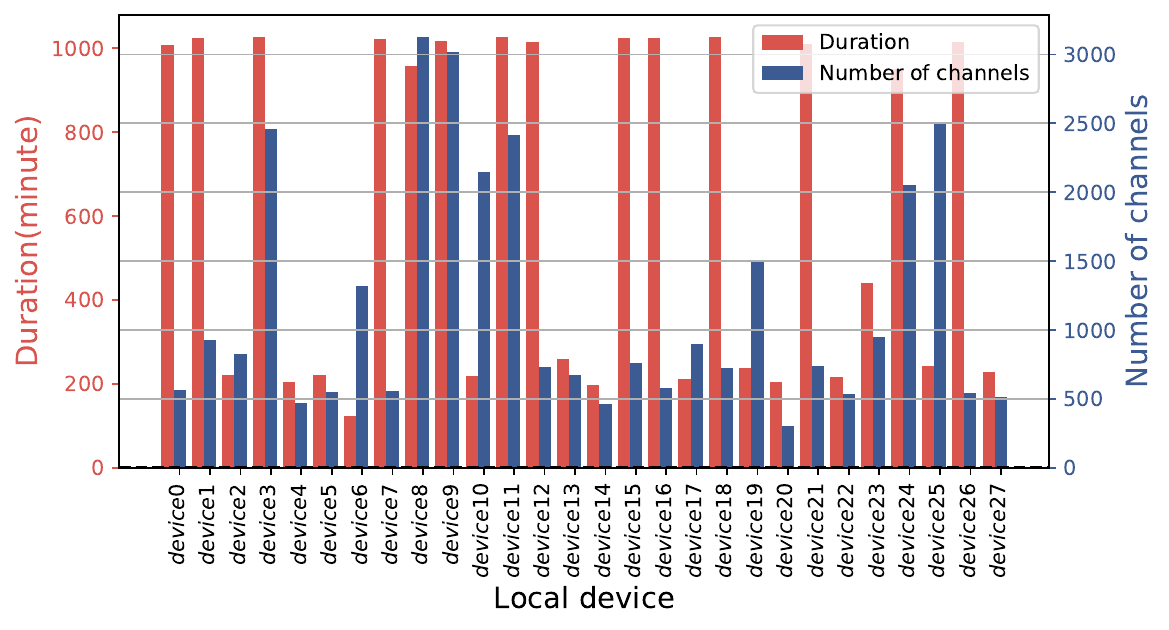}
    \caption{Traffic statistical distribution of the Benign-ALL, including each device's duration and the number of channels.}
    \label{data}
\end{figure}

In the Benign-ALL, the number of IPs communicated by each monitoring device is shown in Fig.~\ref{data}. Each device generates traffic data with at least 302 IP addresses. Benign-ALL includes 32860 channels, in total 45GB.

In addition, we use malware traffic captured by the malware capture facility project~\cite{stratodatasets} over the long term (2013-now) as the malware traffic in our evaluation. This dataset is widely used in the evaluation of malware traffic detection. It covers many types of malware, such as Zeus, Cridex, Emotet, and Trickbot. The types of attacks include DoS, brute force guessing, information leakage, scanning, and C\&C communication.

In this paper, we remove the benign samples from the CTU-ALL dataset since the labels are unreliable. We select the six malware traffic types (Zeus, Emotet, Miuref, Trickbot, Dridex, and Downloadguide) from the malware capture facility project that exhibit the highest number of channels to form the CTU-6 dataset. Meanwhile, we include all other malware traffic from the malware capture facility project to evaluate unknown malware traffic. It comprises 328 malware traffic samples, in total 75781 channel traffic. Table~\ref{dataset} shows the experiment dataset statistics.

To address data imbalance, we employ undersampling across detection tasks to keep the data balanced. Specifically, in known malware traffic detection, we sample benign traffic to 3965, which is the average of the number of six malware channel traffic. In the unknown malware traffic detection, we sample 16430 channel traffic in CTU-6 and CTU-ALL respectively to balance with the benign traffic. To avoid sampling bias, we repeat each experiment set ten times and take the average value as the final evaluation result.

In addition, to ensure that both datasets share similar distributions and to prevent overfitting, we collect traffic from enterprise networks for benign traffic. Meanwhile, we perform traffic preprocessing on every dataset, which includes randomizing IP and MAC addresses and modifying timestamps. Moreover, an analysis of port usage shows a significant overlap between benign and malware traffic. The unique ports make up less than 2\% of benign traffic and 6.93\% and 17.21\% of source and destination ports in malware traffic. Despite some discrepancies resulting from malware design, the overall port usage similarity between the two types of traffic reduces the potential for classification bias.

\subsubsection{Parameter Tuning}
\label{sec:params}
Parameters directly affect the performance of CBSeq. Here, we introduce the selection of main parameters, including (1) the eps of clustering, (2) the dimensionality of embedding, and (3) the maximum sequence length.

\begin{figure}[!t]
    \centering
    \includegraphics[width=0.9\linewidth]{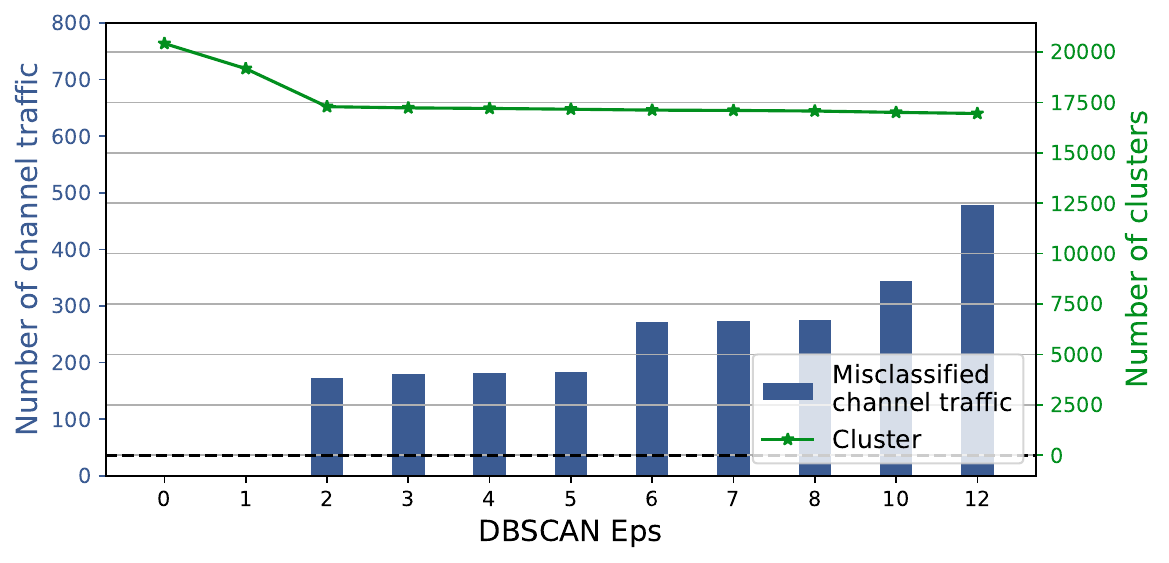}
    \caption{Results on the eps of clustering.}
    \label{cluster}
\end{figure}

\begin{figure*}[!t]
\centering
\subfloat[AUC]{
\begin{minipage}[t]{0.33\textwidth}
\centering
\includegraphics[width=2.2in]{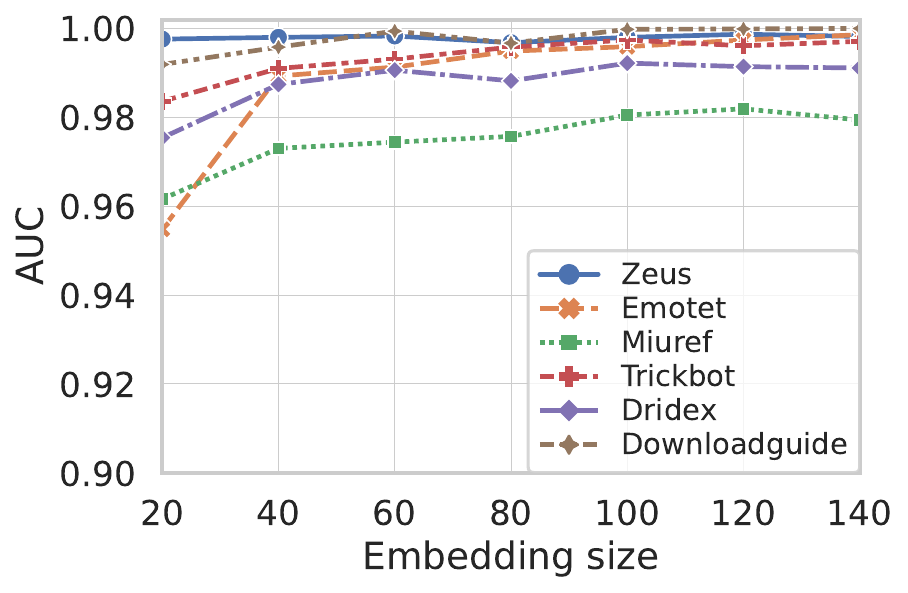}
\end{minipage}%
}%
\subfloat[TPR]{
\begin{minipage}[t]{0.33\textwidth}
\centering
\includegraphics[width=2.1in]{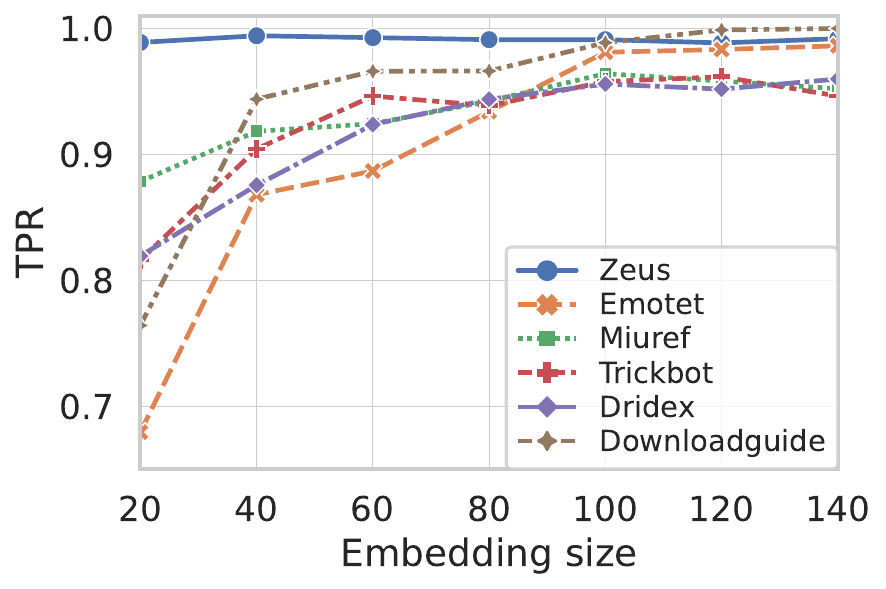}
\end{minipage}%
}%
\subfloat[FPR]{
\begin{minipage}[t]{0.33\textwidth}
\centering
\includegraphics[width=2.1in]{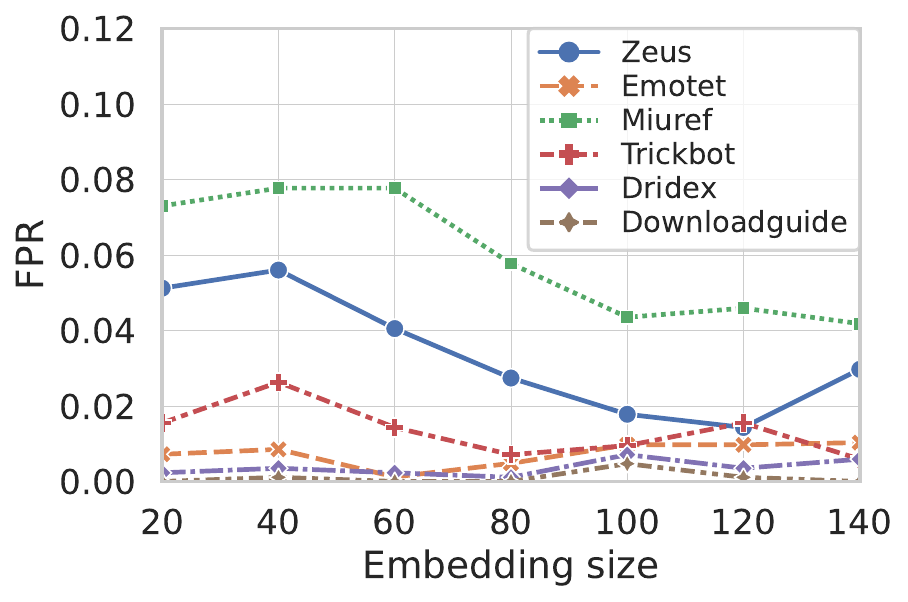}
\end{minipage}%
}%
\caption{Results on the dimensionality of embedding.}
\label{embsize}
\end{figure*}

\begin{figure*}[!t]
\centering
\subfloat[AUC]{
\begin{minipage}[t]{0.33\textwidth}
\centering
\includegraphics[width=2.2in]{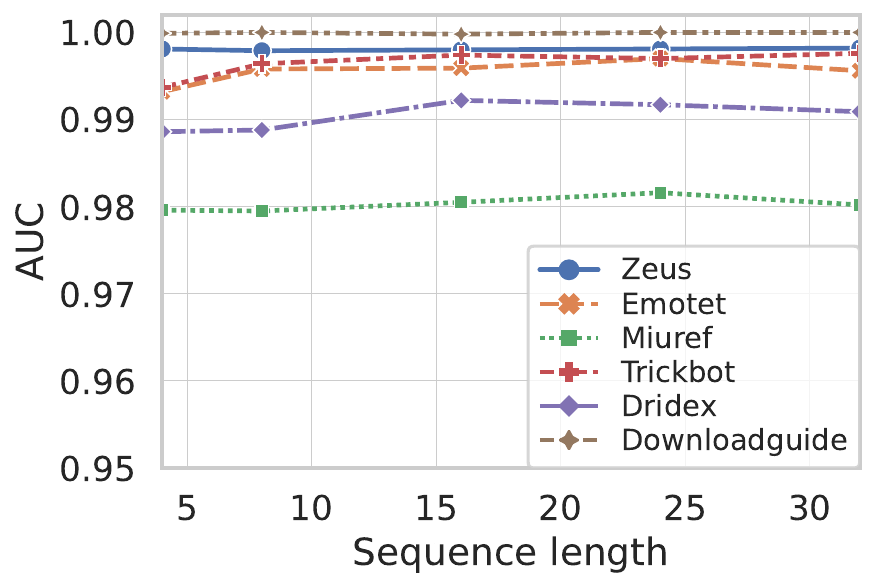}
\end{minipage}%
}%
\subfloat[TPR]{
\begin{minipage}[t]{0.33\textwidth}
\centering
\includegraphics[width=2.1in]{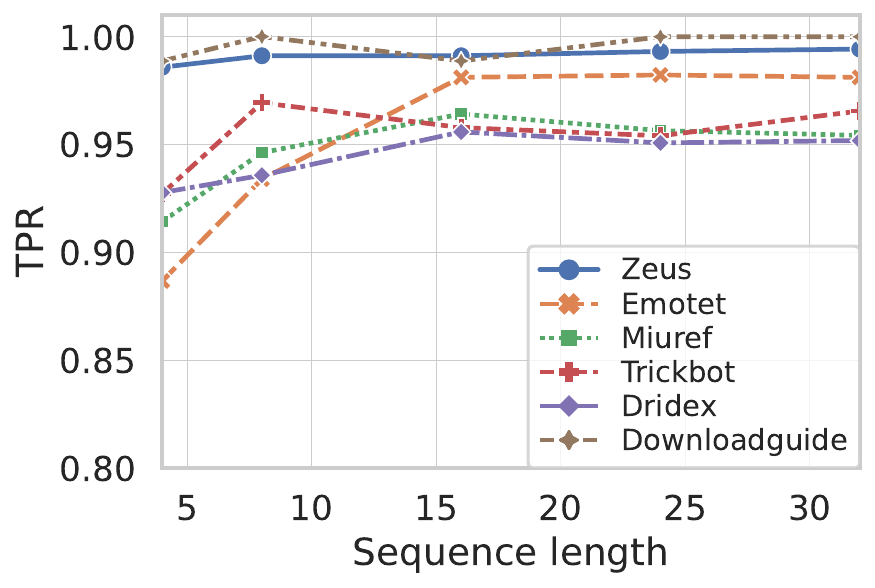}
\end{minipage}%
}%
\subfloat[FPR]{
\begin{minipage}[t]{0.33\textwidth}
\centering
\includegraphics[width=2.1in]{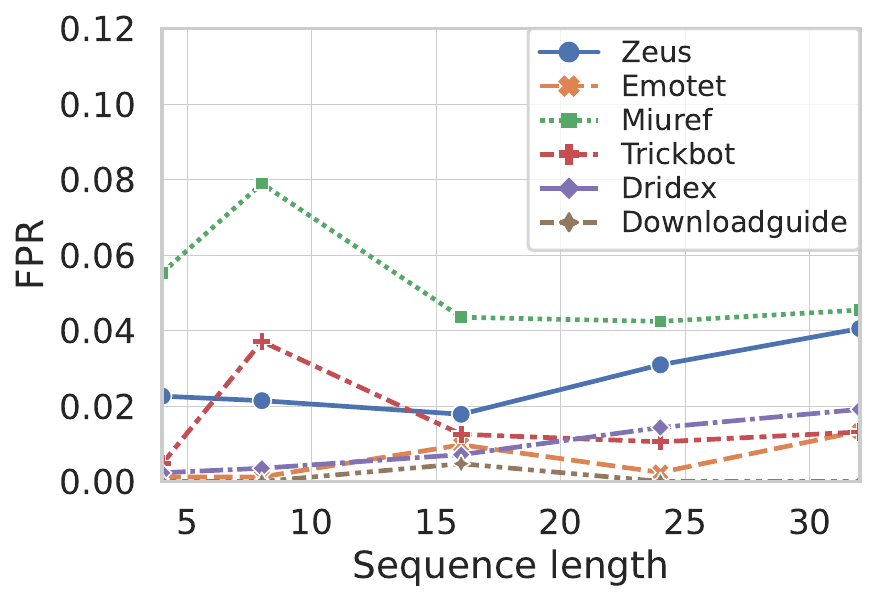}
\end{minipage}%
}%
\caption{Results on the maximum sequence length.}
\label{seqlen}
\end{figure*}

\begin{itemize}
    \item The eps of clustering: We cluster channels with similar activities together by DBSCAN, and the channels with the same cluster share the same label, whether benign traffic or malware traffic. Our goal is to aggregate as much similar channel traffic as possible while guaranteeing that the traffic within each cluster shares the same label. Therefore, we vary the eps parameter to observe the clustering results, as illustrated in Fig.~\ref{cluster}. We observe that when the eps is set to 1, channels within the same cluster possess a single label. However, we observe some channel traffic with misclassified labels when setting a larger eps. For instance, when eps is set to 2, the DBSCAN cluster channel traffic into 17289 clusters. However, 173 instances of channel traffic are misclassified. Therefore, to ensure that each cluster comprises channels with the same label, we set the eps parameter for clustering to 1.
    \item The dimensionality of embedding: The dimensionality of the embedding will directly affect the effectiveness and efficiency of the detector. If the dimensionality is too low, the representation capability of the word vector is insufficient. If the dimensionality is too high, it will easily cause overfitting and affect the detection efficiency. For this reason, we train word vectors with different dimensions and evaluate them for malware traffic detection. The experimental results are shown in Fig.~\ref{embsize}. When the dimension is 100, the detection of each malware traffic can satisfy high and stable performance. Therefore, we choose 100 as the dimension of the word vector.
    \item The maximum sequence length: MSFormer can train the entire sequence to obtain a model for malware traffic detection. However, if the entire sequence is used, it needs to be filled to the maximum sequence length for shorter sequences, which can affect the efficiency of model training and detection. For this reason, we set different maximum sequence lengths to evaluate the effect of sequence length on malware detection. The experimental results are shown in Fig.~\ref{seqlen}. It performs accurately and efficiently when the sequence length is set to 16. Therefore, we set the maximum sequence length to 16.
\end{itemize}

\subsubsection{Setting of the CBSeq}
We use a grid search to determine the parameters used in CBSeq. Specifically, in clustering, we set the time window to 4 hours, eps of DBSCAN to 1, and minPts to 1. In embedding, we set the dimension of embedding to 100. In MSFormer, we set the number of blocks to 6 and the number of headers in multi-head attention to 8. In addition, we use the Adam algorithm with a batch size of 8 to train MSFormer. Moreover, the learning rate is set to 0.00001, and the learning epoch is set to 20. Our model is implemented using PyTorch.

\subsubsection{Beseline Methods}
In order to evaluate and compare the performance of CBSeq, we summarize the following baseline methods:

\begin{itemize}
    \item{Joy and Joy-enhanced~\cite{DBLP:conf/kdd/AndersonM17,DBLP:conf/icnp/McGrewA16}: Joy creates flow features based on statistical analysis, byte distribution matrix, markov chain of packet length/packet time. And then, it uses an RF to build a classifier for malware detection. Joy-enhanced extends the features of Joy by adding TLS ClientHello cipher suite and extension features.}
    \item{CIC-Flow: According to the statistical-based studies~\cite{DBLP:journals/sensors/VegaCHL20,9443025,Vu2022}, CIC-Flow extracts flow statistical features related to time~\cite{DBLP:conf/icissp/LashkariDMG17,DBLP:conf/icissp/Draper-GilLMG16} by CICFlowmeter~\cite{lashkari2017cicflowmeter}, such as time interval, number of packets, duration, TCP flags, and uses an RF to build a classifier for detection.}
    \item{DANTE~\cite{DBLP:conf/esorics/CohenMKMEPS20}: DANTE extracts the destination port sequence of the channel, performs sequence embedding using word2vec and builds an RF classifier for detection.}
    \item {TCC~\cite{fang2021communication}: TCC extracts distribution features, TLS plaintext information, and statistical features from the channel traffic. These features are then processed using a genetic algorithm for extraction, and an RF algorithm is used for malware traffic detection.}
\end{itemize}

\subsubsection{Metrics}
We evaluate all the methods based on the true positive rate (TPR) and false positive rate (FPR). The definitions are as follows:
\begin{equation}
\label{equ1}
\begin{split}
TPR &=\frac{TP}{TP+FN}\\
FPR &=\frac{FP}{FP+TN}
\end{split}
\end{equation}


\begin{table*}[!t]
\renewcommand\arraystretch{1.3}
\setlength\tabcolsep{3px}
\centering
\caption{Result on known malware traffic detection.}
\begin{tabular}{c|c|c|c|c|c|c|c|c|c|c|c|c|c|c|c|c|c|c}
\hline
 & \multicolumn{3}{c|}{Joy} &\multicolumn{3}{c|}{Joy-enhanced} & \multicolumn{3}{c|}{CIC-Flow}& \multicolumn{3}{c|}{DANTE}& \multicolumn{3}{c|}{TCC}& \multicolumn{3}{c}{CBSeq}\\
\cline{2-19}
Metrics&TPR&FPR&AUC&TPR&FPR&AUC&TPR&FPR&AUC&TPR&FPR&AUC&TPR&FPR&AUC&TPR&FPR&AUC\\
\hline
Zeus &0.994	&0.054	&0.983	&\textbf{0.997}	&0.044	&0.982	&0.997	&0.021	&0.991	&0.969	&0.434	&0.769	
&0.982 &	0.016 &	0.991 &0.991	&\textbf{0.018}	&\textbf{0.998}	\\
\hline
Emotet &0.880	&0.006	&0.961	&0.959	&0.005&0.981	&0.889 &	0.002	&0.957	&0.896	&0.062	&0.920
&0.904&	0.015&	0.988&\textbf{0.981}	&\textbf{0.010}	&\textbf{0.996}\\
\hline
Miuref	&0.964	&0.056	&0.967	&0.964&0.049&0.970	&\textbf{0.988}&\textbf{0.041}	&0.978	&0.779	&0.079 &0.862	
&0.918&	0.053&	0.971&0.964 &0.044 &\textbf{0.981}\\
\hline
Trickbot&0.944	&0.027	&0.968	&0.954	&0.027	&0.973	&\textbf{0.978}&	0.017	&0.988	&0.897	&0.050	&0.942	
&0.926&	0.011&	0.963&0.958	&\textbf{0.010}	&\textbf{0.997}\\
\hline
Dridex	&0.902	&0.016	&0.964	&0.917	&0.017	&0.963	&\textbf{0.973}&	0.025	&0.980	&0.932	&0.064	&0.947	&0.937&	0.011&	0.989&0.956	&\textbf{0.007}	&\textbf{0.992}\\
\hline
Downloadguide	&0.920	&0.003	&0.959	&0.868	&0.002	&0.946&	0.972	&0.021	&0.984	&0.989	&0.048	&0.992	&0.891&	0.010&	0.973&\textbf{0.989}	&\textbf{0.005} &\textbf{0.999}\\
\hline
\hline
\textbf{Average}	&0.934	&0.027	&0.967	&0.943	&0.024	&0.969&	0.966	&0.021	&0.980	&0.910	&0.123	&0.905
&0.926	&0.019&	0.979&\textbf{0.973}	&\textbf{0.016} &\textbf{0.994}\\
\hline

\multicolumn{16}{l}{*\textbf{Bold} indicates the best value compared to other baseline methods.}\\
\end{tabular}
\label{tab1}
\end{table*}

In addition, we use ROC and AUC to evaluate the general performance of the detector. AUC is the probability that a randomly-chosen positive example is ranked more highly than a randomly-chosen negative example. An AUC close to 1 indicates a more effective detector, whereas an AUC of 0.5 implies random identification by the detector.

\subsection{Encrypted Known Malware Traffic Detection}

In this section, we evaluate the CBSeq's performance under encrypted known malware traffic detection, including six malware. Specifically, for each type of malware, we train a binary detector using the CTU-6 dataset for malware traffic and the Benign-ALL dataset for benign traffic, which allows us to analyze and evaluate CBSeq's ability to detect different malware. Moreover, we utilize the same benign traffic to compare the detection capability across each malware type. We also employ 5-fold cross-validation to ensure more reliable results.

To fully explore the contribution of CBSeq, we utilize the five baseline methods mentioned above for comparison. Note that we use RF as the classifier for the baseline methods since RF exhibits better detection capabilities than other machine learning classifiers.

In this experiment, we evaluate the detection performance of CBSeq for different malware traffic by measuring TPR, FPR, and AUC. Table~\ref{tab1} illustrates the results. We find that CBSeq can perform better in detecting all six types of malware traffic, with their AUC reaching between 0.981 and 0.999. In all the detections, the TPR values are above 0.95, and the FPR values are below 0.05. Moreover, the detection performance of CBSeq is closely related to the network behavior of the malware. For example, Miuref, which has the highest FPR in our results, involves advertisement click fraud in its attack process. Miuref simulates user behavior on infected devices, clicking on advertisements to generate revenue from advertisers. This behavior closely resembles benign traffic, leading to a higher FPR. In general, CBSeq is effective at detecting malware traffic. Compared to other baseline methods, CBSeq achieves higher AUC for all malware traffic, lower FPR for five malware traffic, and higher TPR for two malware traffic.

In terms of Joy and Joy-enhanced, Joy achieves a TPR of above 0.95 for only two malware and an FPR of below 0.05 for four malware. Joy does not achieve a TPR above 0.95 and an FPR below 0.05 for any malware. Joy-enhanced achieves a TPR above 0.95 for four types of malware traffic and an FPR below 0.05 for six types of malware traffic. We observe that Joy-enhanced has better detection capability than Joy. However, Joy-enhanced does not have a more significant improvement because our dataset contains a large amount of other protocol traffic in addition to TLS traffic, especially malware traffic, which only accounts for a minority of TLS connections. Therefore, Joy-enhanced is unable to extract TLS fingerprint information from traffic using other protocols, such as HTTP.

In terms of CIC-Flow, we find that it outperforms Joy and Joy-enhanced. In CIC-Flow detection, only Emotet has a TPR below 0.95, while all other malware meets more than 0.95 of TPR and less than 0.05 of FPR. In addition, Zeus, Miuref, Trickbot, and Dridex achieve the highest TPR, and Miuref reaches the lowest FPR in comparing all methods.

In terms of DANTE, only Downloadguide's detection capability meets more than 0.95 of TPR and less than 0.05 of FPR. The detection FPR for Zeus traffic exceeds 0.434, and the TPRs for three types of malware traffic are below 0.9. Therefore, the destination port sequence only achieves effective detection for specific malware. It cannot effectively detect general malware traffic.

In terms of TCC, out of the six types of malware traffic, only Zeus achieves a TPR exceeding 0.95. Aside from Miuref, the FPRs for detecting the other five types of malware traffic are less than 0.05. Therefore, only Zeus could satisfy the TPR above 0.95 and FPR below 0.05. Moreover, similar to Joy-enhanced, TCC also uses plaintext content from the TLS as a feature, which hinders these methods from maintaining high accuracy rates in detecting traffic from other protocols.

The experimental results conclusively demonstrate the effectiveness of CBSeq in encrypted known malware traffic detection.

\begin{figure*}[!t]
\centering
\subfloat[Joy-enhanced]{
\begin{minipage}[t]{0.25\linewidth}
\includegraphics[width=1.5in]{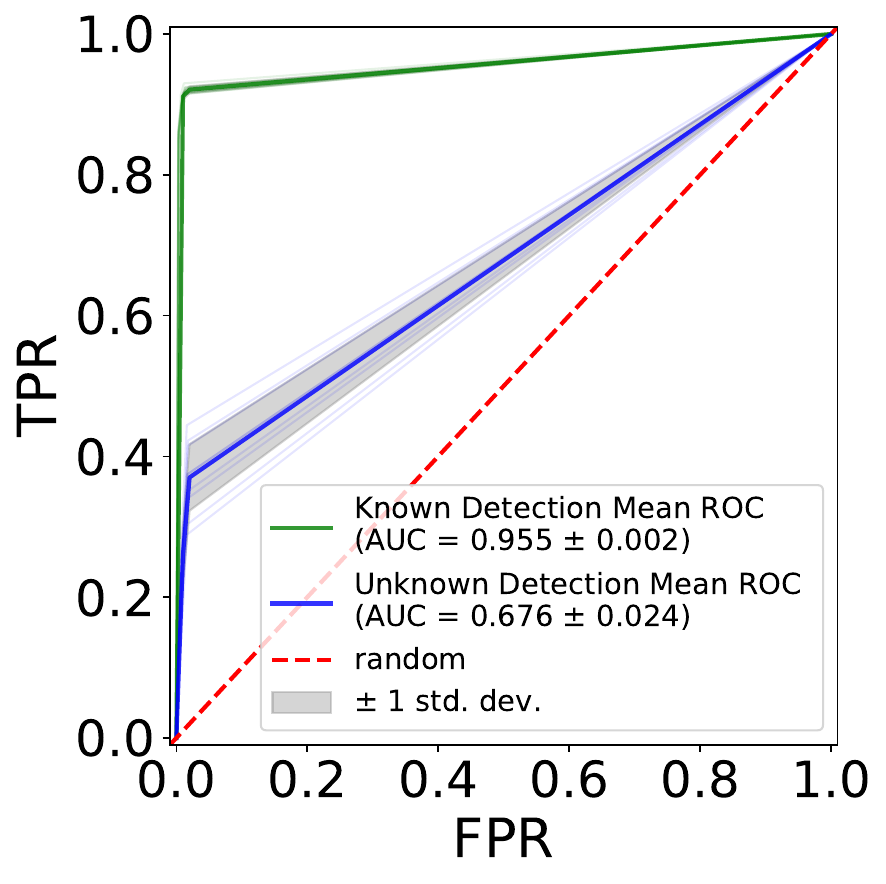}
\end{minipage}%
}%
\subfloat[CIC-Flow]{
\begin{minipage}[t]{0.25\linewidth}
\includegraphics[width=1.5in]{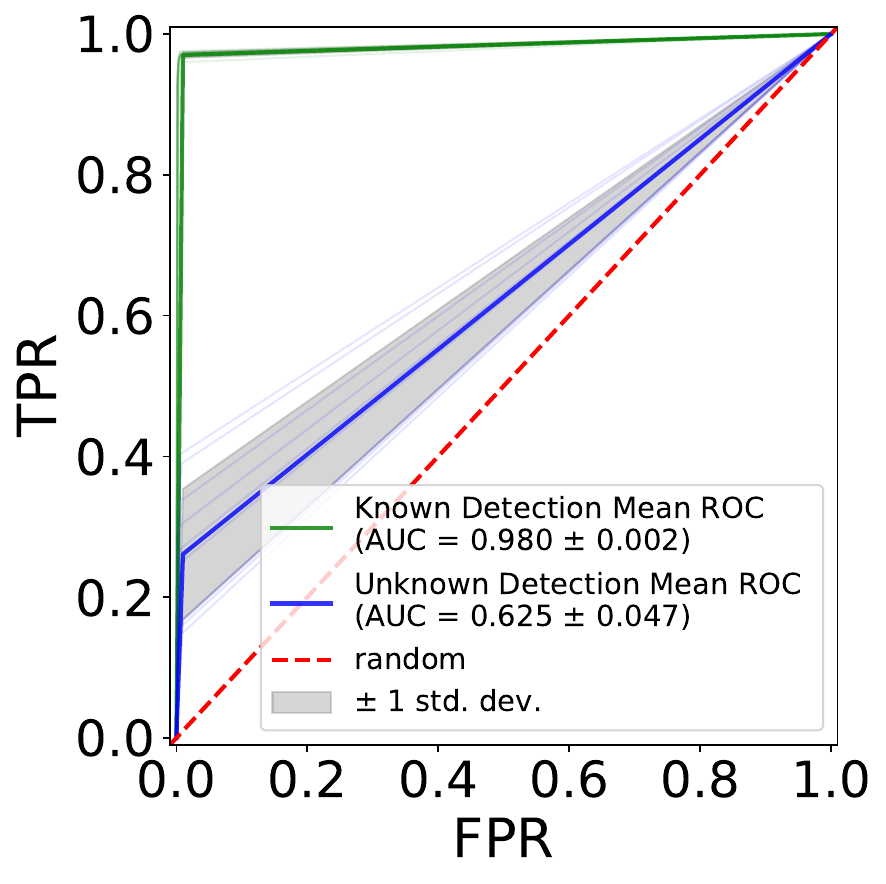}
\end{minipage}%
}%
\subfloat[TCC]{
\begin{minipage}[t]{0.25\linewidth}
\includegraphics[width=1.5in]{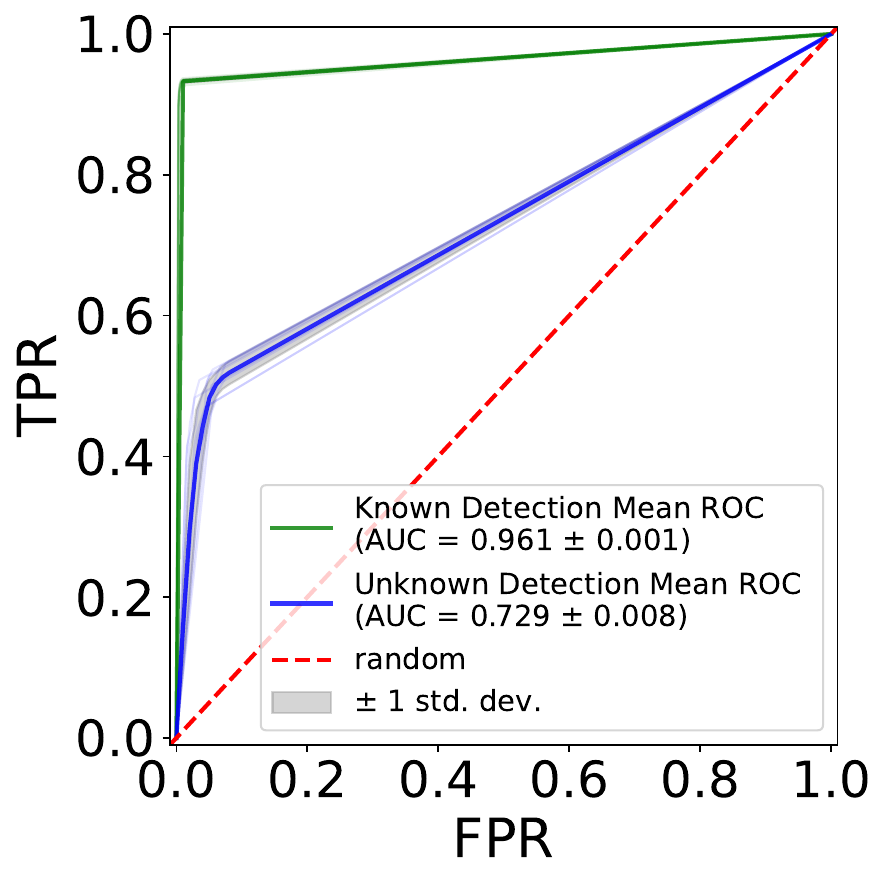}
\end{minipage}%
}%
\subfloat[CBSeq]{
\begin{minipage}[t]{0.25\linewidth}
\includegraphics[width=1.5in]{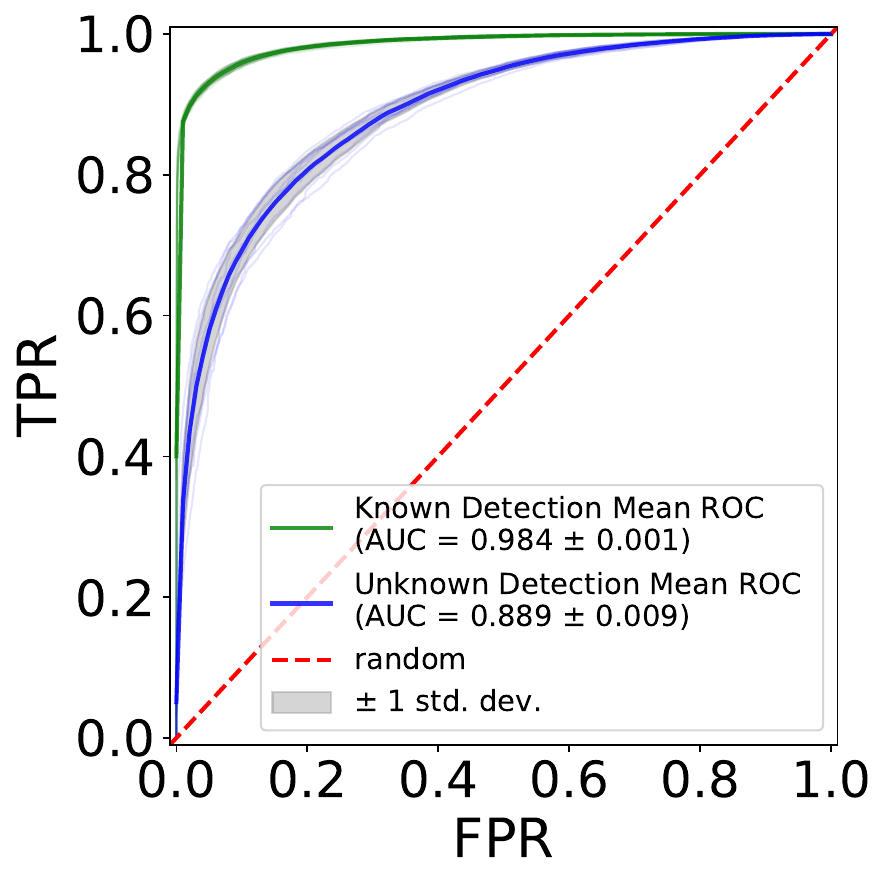}
\end{minipage}%
}%
\caption{The ROC curves and the AUC values of CBSeq and baseline methods under unknown malware traffic detection.}
\label{unknown}
\end{figure*}

\subsection{Encrypted Unknown Malware Traffic Detection}
In this section, we evaluate the performance of CBSeq in encrypted unknown malware traffic detection by detecting malware traffic in CTU-ALL. We train a binary classifier using the CTU-6 dataset and 50\% of Benign-ALL dataset and utilize this detector to identify malware traffic in the CTU-ALL dataset and benign traffic in another 50\% of Benign-ALL dataset. Note that the detector's task is to distinguish whether the traffic is malicious or not, rather than identifying the specific type of malware.

Meanwhile, we train a binary classifier for detecting known malware traffic, utilizing the CTU-6 and Benign-ALL datasets. The detection results from this process are considered the theoretical upper limit for detecting unknown malware traffic. In order to more closely match the actual environment, when dividing the training and test sets in Benign-ALL, the traffic of the same channel can only exist in the training set or only in the test set.

To evaluate CBSeq's performance, we compare it to Joy-enhanced, CIC-Flow, and TCC, which are known to be more effective at detecting known malware traffic. Fig.~\ref{unknown} shows the ROC curves and AUC values for their detection. Since our experiments could not be applied to cross-validation, we randomly sample the test set and repeat it ten times to obtain more reliable detection results. 

CBSeq achieves an AUC value of 0.984 for known malware traffic detection and 0.889 for unknown malware traffic detection. It indicates that CBSeq achieves effective detection in closed known malware traffic detection and maintains the AUC value above 0.88 in unknown malware traffic detection. Compared with the baseline methods, CBSeq improves the AUC value by 0.160 for unknown malware traffic detection.

For Joy-enhanced, we observe that it has an AUC value of 0.955 for known detection. However, there is a large amount of unknown malware traffic in the real world that is not present in the training set, which will result in lower performance of the detector. With unknown malware detection, the Joy-enhanced detector only achieves an AUC of 0.676.

In terms of CIC-Flow, it can achieve the AUC value of 0.980 for known malware detection while only 0.625 for unknown malware detection. It can be seen that the features extracted by Joy-enhanced and CIC are only applicable to known malware detection in the closed environment. They cannot effectively characterize unknown malware traffic.

In terms of TCC, it achieves an AUC of 0.961 for known malware traffic detection and 0.729 for unknown malware traffic detection. Moreover, we observe that TCC conducts statistical analysis at the channel granularity, which performs better than Joy-enhanced and CIC at the traditional flow granularity. However, CBSeq has a higher AUC than TCC by 0.160. It validates that the behavior sequence proposed by CBSeq in this paper is more stable in unknown malware traffic detection.

The experimental results conclusively demonstrate the effectiveness of CBSeq in encrypted unknown malware detection.

\begin{figure}[!t]
    \centering
    \includegraphics[width=0.9\linewidth]{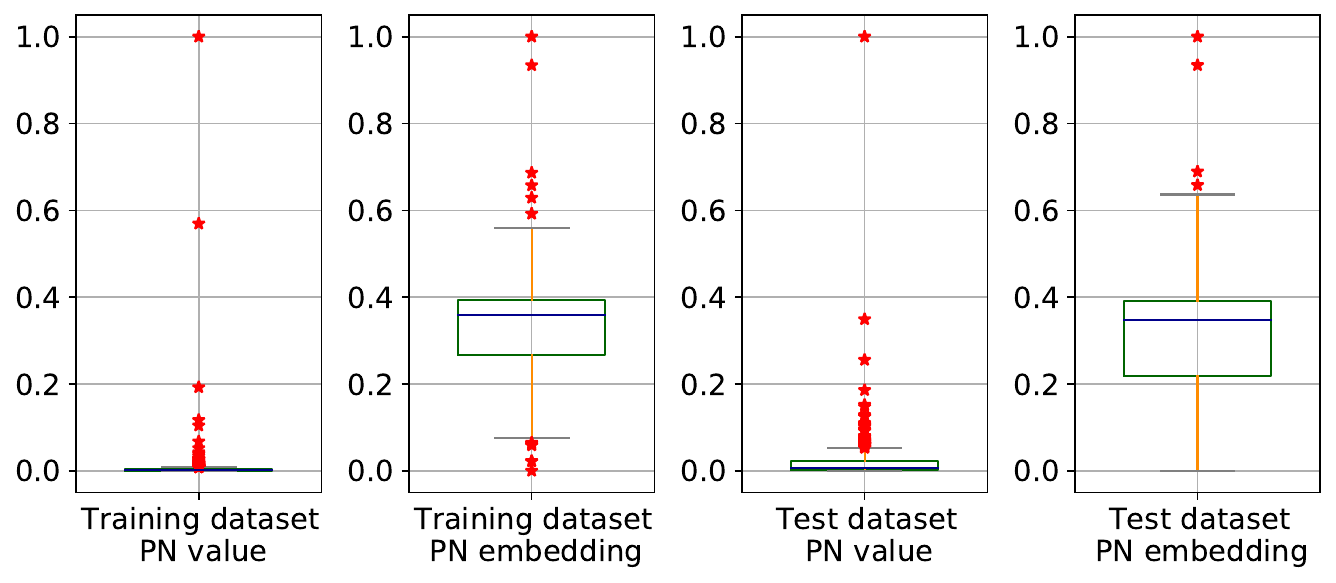}
    \caption{The comparison results of embedding in PN sequence.}
    \label{box_pn}
\end{figure}
\begin{figure}[!t]
    \centering
    \includegraphics[width=0.9\linewidth]{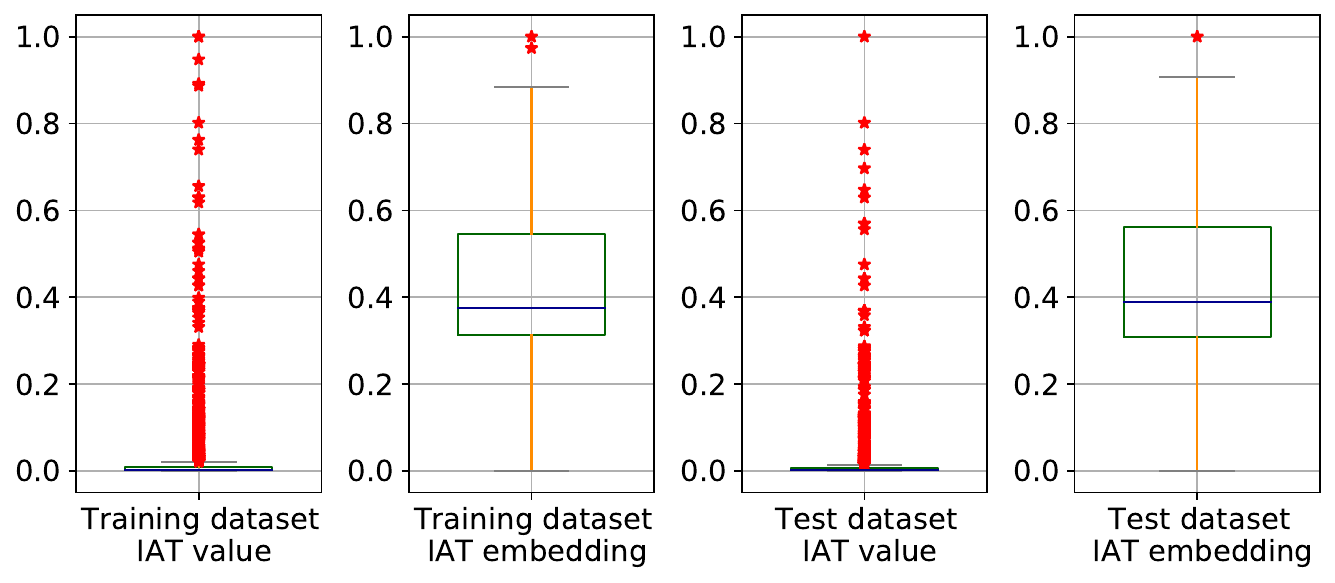}
    \caption{The comparison results of embedding in IAT sequence.}
    \label{box_iat}
\end{figure}

\subsection{Analysis on Embedding}

In this section, we analyze the behavior sequence embedding proposed by CBSeq, which employs word2vec to convert the original behavior sequence. To train the word2vec model, we utilize a variety of long-term collected public traffic datasets. These datasets include the WIDE Project backbone traffic (from 2013 to now)~\cite{271335}, malware-traffic-analysis.net (from 2013 to now)~\cite{malware-traffic-analysis}. These datasets contain authentic and diverse network activities and malicious attacks, providing a large-scale and diversified traffic pool. Moreover, to ensure comprehensive training and representation of various types of traffic, we implement the undersampling for data balancing, obtaining improved results.

\begin{figure}[!t]
\centering
\subfloat[Global]{
\begin{minipage}[t]{0.5\linewidth}
\includegraphics[width=1.5in]{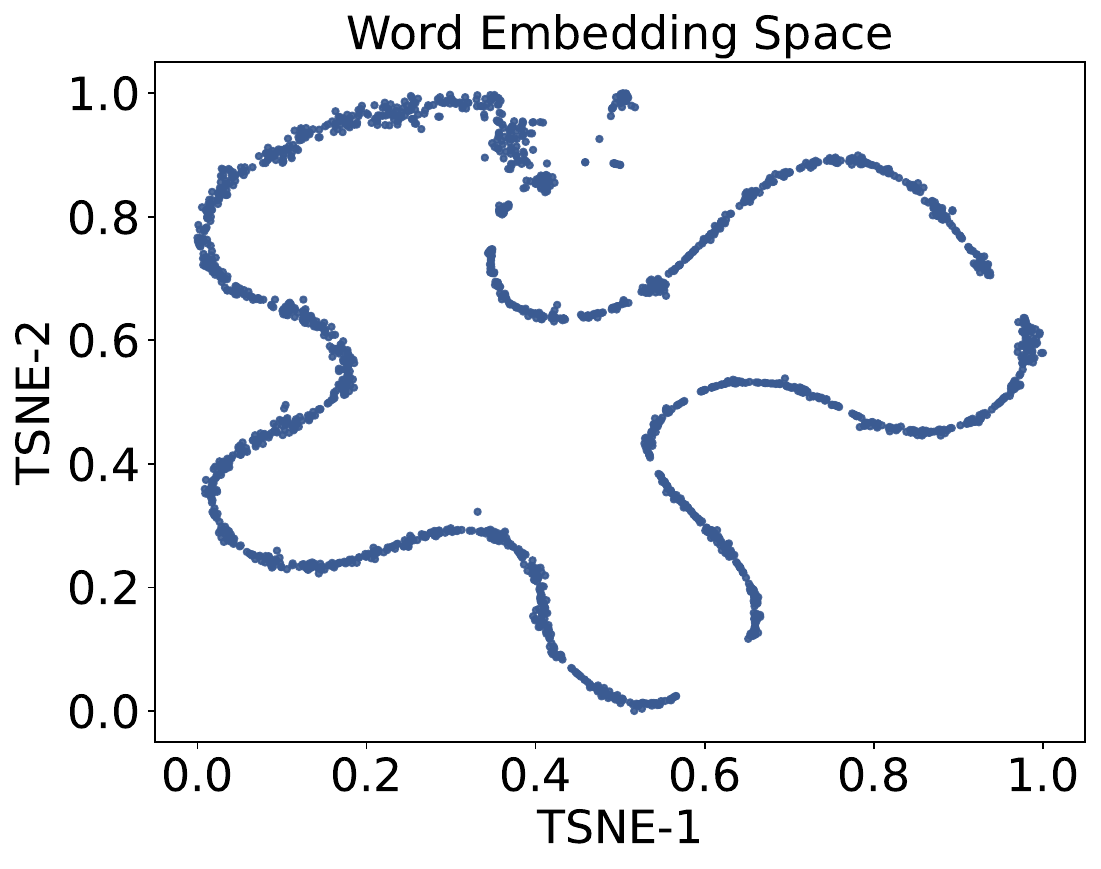}
\end{minipage}%
}%
\subfloat[Local]{
\begin{minipage}[t]{0.5\linewidth}
\includegraphics[width=1.6in]{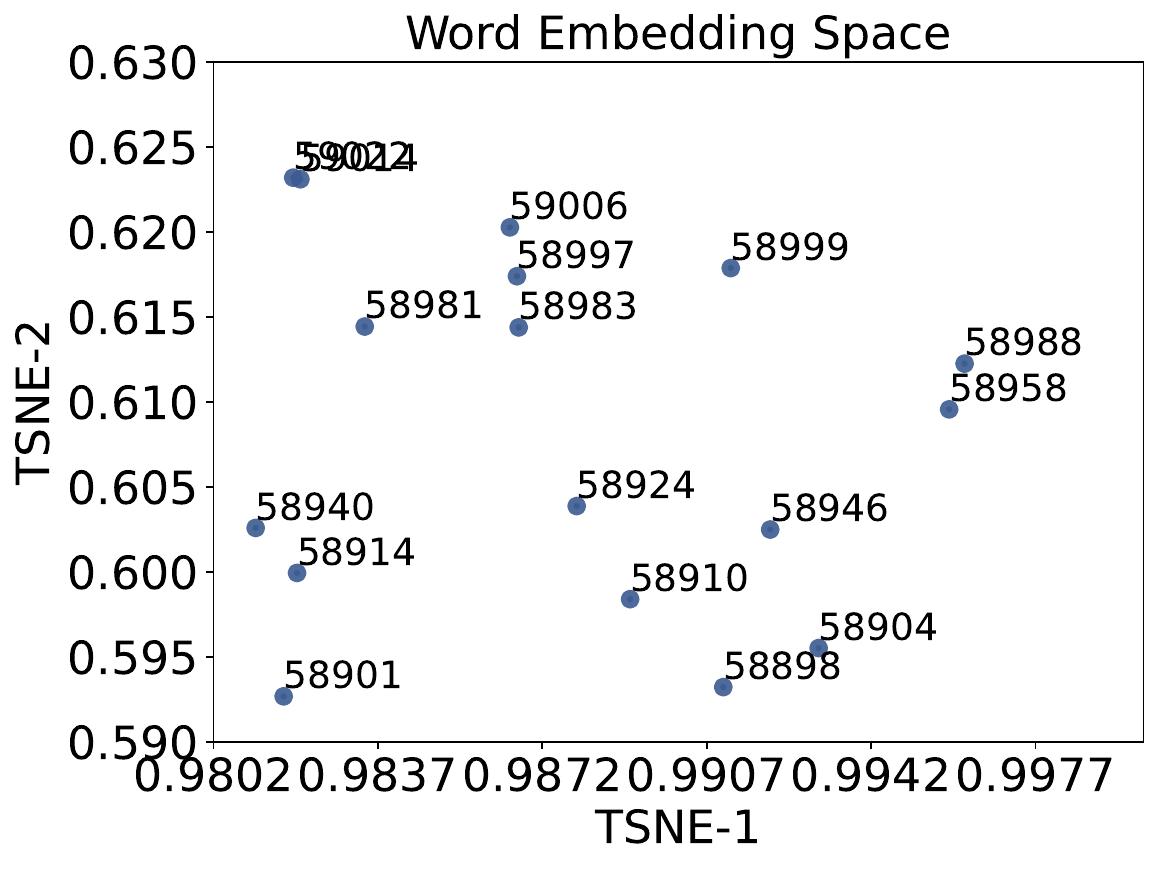}
\end{minipage}%
}%
\caption{The TSNE visualization of embedding in SP sequence. The picture on the right is a local enlargement on the left.}
\label{sca_sp}
\end{figure}

\begin{figure}[!t]
\subfloat[Global]{
\begin{minipage}[t]{0.5\linewidth}
\centering
\includegraphics[width=1.5in]{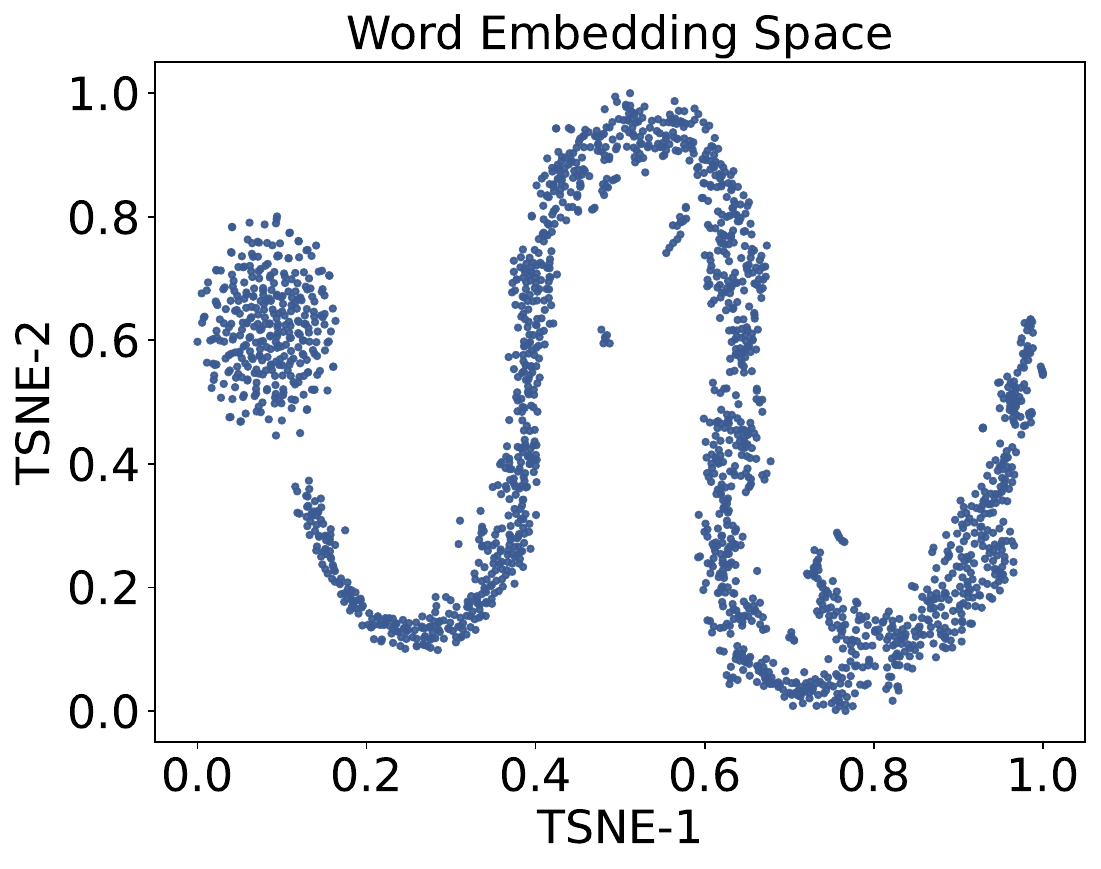}
\end{minipage}%
}%
\subfloat[Local]{
\begin{minipage}[t]{0.5\linewidth}
\includegraphics[width=1.6in]{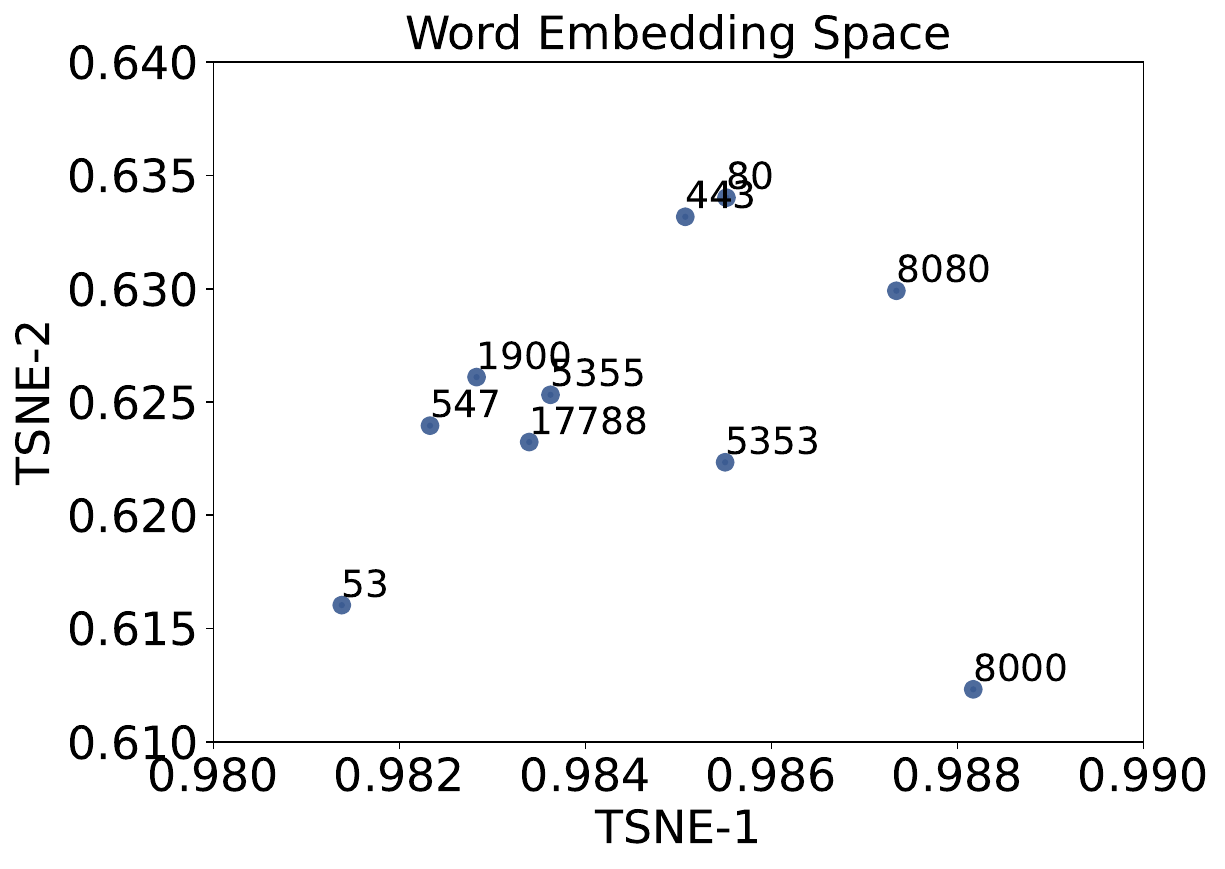}
\end{minipage}%
}%
\caption{The TSNE visualization of embedding in DP sequence. The picture on the right is a local enlargement of the left.}
\label{sca_dp}
\end{figure}

Fig.~\ref{box_pn} and Fig.~\ref{box_iat} show the comparison of the original values and embeddings within the PN sequence and IAT sequence. We first apply the PCA algorithm to reduce both the original values and embeddings to one dimension, followed by a box plot to visualize the data distribution. We observe that there are more outliers in the PN value and IAT value, and the distribution range of the test dataset and training dataset is inconsistent, which will directly affect the detection effect of the model. The embedding process leads to the training and test datasets having the same distribution, and it significantly reduces the number of outliers. Therefore, embedding can solve the problem of large data differentiation.

Fig.~\ref{sca_sp} and Fig.~\ref{sca_dp} show the TSNE visualization of the port embedding within the SP sequence and the DP sequence. We observe that the embedding distributions of both SP and DP are uniform, and there are no outliers. In SP embedding, ports with similar values are embedded at similar distances because the attacking channels usually access the specified server through multiple continuous ports. Moreover, port numbers with the same service are closer in DP embedding. For example, ports 443, 80, and 8080 are usually found in web services, so they are closer in embedding space. Thus, embedding can correlate behavior connections between port numbers to characterize the channel's activity behavior.

Through our analysis of the embedding, we observe that it effectively addresses the issue of large data differentiation and can correlate sequence behavior intent.

\subsection{Analysis on Behavior Sequence}

\newcommand{\tabincell}[2]{\begin{tabular}{@{}#1@{}}#2\end{tabular}}  
\begin{table*}[!t]
\renewcommand\arraystretch{1.3}
\setlength\tabcolsep{3px}
\centering
\caption{Result of Analysis on behavior sequence.}
\begin{tabular}{c|c|c|c|c|c|c|c|c|c|c|c|c|c|c|c|c|c|c}
\hline
 & \multicolumn{3}{c|}{Zeus} &\multicolumn{3}{c|}{Emotet} & \multicolumn{3}{c|}{Miuref}& \multicolumn{3}{c|}{Trickbot}& \multicolumn{3}{c|}{Dridex}&\multicolumn{3}{c}{Downloadguide}\\
\cline{2-19}
Metrics&TPR&FPR&AUC&TPR&FPR&AUC&TPR&FPR&AUC&TPR&FPR&AUC&TPR&FPR&AUC&TPR&FPR&AUC\\
\hline
PN Sequence&	\textbf{0.996}&	0.479&	0.872&	0.330&	\textbf{0.001}&	0.951&	0.741&	0.077&	0.925&	0.460&	\textbf{0.0}&	0.948&	0.863&	\textbf{0.0}&	0.975&	0.966&	\textbf{0.0}&	\textbf{0.999}\\
\hline
IAT Sequence&	\textbf{0.970}&	0.131&	0.956&	0.708&	\textbf{0.034}&	0.935&	0.829&	0.185&	0.871&	0.383&	\textbf{0.040}&	0.873&	0.643&	0.107&	0.866&	0.315&	\textbf{0.019}&	0.868\\
\hline
SP Sequence&	\textbf{0.979}&	\textbf{0.049}&	\textbf{0.994}&	0.689&	\textbf{0.013}&	0.976&	0.596&	0.114&	0.859&	0.732&	\textbf{0.019}&	0.941&	0.751&	\textbf{0.014}&	0.959&	0.730&	\textbf{0.0}&	0.961\\
\hline
DP Sequence &	\textbf{0.979}&	0.676&	0.764&	0.774&	\textbf{0.044}&	0.898&	0.438&	0.117&	0.707&	0.739&	\textbf{0.046}&	0.895&	0.651&	0.096&	0.834&	0.337&	\textbf{0.012}&	0.806\\
\hline
No PN Sequence&	0.994&	0.041&	0.998&	\underline{0.943}&	0.012&	0.989&	\underline{0.849}&	\underline{0.097}&	\underline{0.951}&	\underline{0.920}&	\underline{0.073}&	\underline{0.979}&	\underline{0.928}&	0.026&	0.984&	\underline{0.910}&	0.004&	0.997\\
\hline
No IAT Sequence&	0.989&	0.044&	0.998&	0.962&	0.005&	0.996&	\underline{0.920}&	\underline{0.154}&	\underline{0.959}&	0.973&	0.035&	0.997&	\underline{0.948}&	0.018&	0.988&	0.978&	0.0&	 1.0\\
\hline
No SP Sequence&	0.988&	\underline{0.099}&	0.983&	\underline{0.877}&	0.023&	0.989&	\underline{0.922}&	\underline{0.074}&	\underline{0.968}&	\underline{0.908}&	0.032&	0.985&	\underline{0.908}&	0.010&	0.984&	0.966&	0.0&	0.999\\
\hline
No DP Sequence &	0.988&	0.027&	0.997&	\underline{0.906}&	0.006&	0.994&	0.962&	\underline{0.098}&	0.980&	0.950&	0.020&	0.995&	\underline{0.944}&	0.006&	0.991&	1.0&	0.0&	1.0\\
\hline

\multicolumn{19}{l}{\tabincell{l}{*\textbf{Bold} denotes values that meet TPR$>$0.95, FPR$<$0.05, and AUC$>$0.98 in PN/IAT/SP/DP Sequence.\\ *\underline{Underline} denotes values that don't meet TPR$>$0.95, FPR$<$0.05 and AUC$>$0.98 in No PN/IAT/SP/DP Sequence.}}\\
\end{tabular}
\label{behavior}
\end{table*}

In this section, we analyze four sequences proposed in CBSeq: PN sequence, IAT sequence, SP sequence, and DP sequence. First, we evaluate the importance of each sequence for malware detection using a single sequence. Second, by removing one sequence at a time, we evaluate the interactions between the remaining three sequences. The experimental results are shown in Table~\ref{behavior}.

We observe that each of the four sequences characterizes malware traffic at different levels. They show different advantages for different malware. For example, the PN sequence can better characterize the Downloadguide, but it has too high FPR for Zeus and too low TPR for Emotet and Trickbot.

Next, we remove a specific sequence for the behavior sequence. We observe that removing any of the sequences degrades the detection performance to varying degrees. Therefore, there is no redundant relationship among the four sequences. Using the four sequences together to construct the behavior sequence can better mine the attacking intent.

\section{Conclusion}
In this paper, we present a method to detect malware traffic from a large amount of complex benign traffic. The approach provides a behavior sequence and designs a detector for known and unknown malware traffic detection. It is cross-protocol, robust, and capable of discovering unknown attacks. By evaluating CBSeq, it can detect different known malware traffic and unknown malware traffic. Moreover, we compare CBSeq with other baseline methods, demonstrating the best performance.

In future work, we plan to design an algorithm to measure the similarity between channel activities, aggregating behaviorally associated channels based on similarity scores. In addition, we will improve MSFormer by learning the relative positions of elements inside the behavior sequence. Moreover, we aim to extend our work to encompass malware behavior beyond network traffic, including intranet lateral movement, boot survival, and privilege escalation. This can potentially be achieved by integrating an endpoint-deployed IDS with our existing system to detect and analyze malicious activities more comprehensively.


\bibliographystyle{IEEEtran}
\bibliography{mybib}

\begin{thebibliography}{10}
\providecommand{\url}[1]{#1}
\csname url@samestyle\endcsname
\providecommand{\newblock}{\relax}
\providecommand{\bibinfo}[2]{#2}
\providecommand{\BIBentrySTDinterwordspacing}{\spaceskip=0pt\relax}
\providecommand{\BIBentryALTinterwordstretchfactor}{4}
\providecommand{\BIBentryALTinterwordspacing}{\spaceskip=\fontdimen2\font plus
\BIBentryALTinterwordstretchfactor\fontdimen3\font minus
  \fontdimen4\font\relax}
\providecommand{\BIBforeignlanguage}[2]{{%
\expandafter\ifx\csname l@#1\endcsname\relax
\typeout{** WARNING: IEEEtran.bst: No hyphenation pattern has been}%
\typeout{** loaded for the language `#1'. Using the pattern for}%
\typeout{** the default language instead.}%
\else
\language=\csname l@#1\endcsname
\fi
#2}}
\providecommand{\BIBdecl}{\relax}
\BIBdecl

\bibitem{9519384}
X.~Li, B.~A. Azad, A.~Rahmati, and N.~Nikiforakis, ``Good bot, bad bot:
  Characterizing automated browsing activity,'' in \emph{2021 IEEE Symposium on
  Security and Privacy (SP)}, 2021, pp. 1589--1605.

\bibitem{DBLP:journals/tifs/ShenLZDH21}
M.~Shen, Y.~Liu, L.~Zhu, X.~Du, and J.~Hu, ``Fine-grained webpage
  fingerprinting using only packet length information of encrypted traffic,''
  \emph{{IEEE} Trans. Inf. Forensics Secur.}, vol.~16, pp. 2046--2059, 2021.

\bibitem{DBLP:journals/tifs/ShenWZW17}
M.~Shen, M.~Wei, L.~Zhu, and M.~Wang, ``Classification of encrypted traffic
  with second-order markov chains and application attribute bigrams,''
  \emph{{IEEE} Trans. Inf. Forensics Secur.}, vol.~12, no.~8, pp. 1830--1843,
  2017.

\bibitem{9277523}
K.~Shaukat, S.~Luo, V.~Varadharajan, I.~A. Hameed, and M.~Xu, ``A survey on
  machine learning techniques for cyber security in the last decade,''
  \emph{IEEE Access}, vol.~8, pp. 222\,310--222\,354, 2020.

\bibitem{9720753}
M.~E. Zadeh Nojoo~Kambar, A.~Esmaeilzadeh, Y.~Kim, and K.~Taghva, ``A survey on
  mobile malware detection methods using machine learning,'' in \emph{2022 IEEE
  12th Annual Computing and Communication Workshop and Conference (CCWC)},
  2022, pp. 0215--0221.

\bibitem{DBLP:journals/sensors/VegaCHL20}
A.~C. Vega, I.~S. Crespo{-}Mart{\'{\i}}nez, {\'{A}}.~M.~G. Higueras, and C.~F.
  Llamas, ``Flow-data gathering using netflow sensors for fitting
  malicious-traffic detection models,'' \emph{Sensors}, vol.~20, no.~24, p.
  7294, 2020.

\bibitem{9443025}
M.~Gohari, S.~Hashemi, and L.~Abdi, ``Android malware detection and
  classification based on network traffic using deep learning,'' in \emph{2021
  7th International Conference on Web Research (ICWR)}, 2021, pp. 71--77.

\bibitem{Vu2022}
A.-H. Vu, M.-Q. Nguyen-Khac, X.-T. Do, and K.-H. Le, \emph{A Real-Time
  Evaluation Framework For Machine Learning-Based IDS}.\hskip 1em plus 0.5em
  minus 0.4em\relax Cham: Springer International Publishing, 2022, pp.
  317--329.

\bibitem{DBLP:conf/ndss/MirskyDES18}
Y.~Mirsky, T.~Doitshman, Y.~Elovici, and A.~Shabtai, ``Kitsune: An ensemble of
  autoencoders for online network intrusion detection,'' in \emph{25th Annual
  Network and Distributed System Security Symposium, {NDSS} 2018, San Diego,
  California, USA, February 18-21, 2018}.\hskip 1em plus 0.5em minus
  0.4em\relax The Internet Society, 2018.

\bibitem{DBLP:conf/kdd/AndersonM17}
B.~Anderson and D.~A. McGrew, ``Machine learning for encrypted malware traffic
  classification: Accounting for noisy labels and non-stationarity,'' in
  \emph{Proceedings of the 23rd {ACM} {SIGKDD} International Conference on
  Knowledge Discovery and Data Mining, Halifax, NS, Canada, August 13 - 17,
  2017}.\hskip 1em plus 0.5em minus 0.4em\relax {ACM}, 2017, pp. 1723--1732.

\bibitem{DBLP:conf/icnp/McGrewA16}
D.~A. McGrew and B.~Anderson, ``Enhanced telemetry for encrypted threat
  analytics,'' in \emph{24th {IEEE} International Conference on Network
  Protocols, {ICNP} 2016, Singapore, November 8-11, 2016}.\hskip 1em plus 0.5em
  minus 0.4em\relax {IEEE} Computer Society, 2016, pp. 1--6.

\bibitem{DBLP:conf/esorics/CohenMKMEPS20}
D.~Cohen, Y.~Mirsky, M.~Kamp, T.~Martin, Y.~Elovici, R.~Puzis, and A.~Shabtai,
  ``{DANTE:} {A} framework for mining and monitoring darknet traffic,'' in
  \emph{Computer Security - {ESORICS} 2020 - 25th European Symposium on
  Research in Computer Security, {ESORICS} 2020, Guildford, UK, September
  14-18, 2020, Proceedings, Part {I}}, ser. Lecture Notes in Computer Science,
  L.~Chen, N.~Li, K.~Liang, and S.~A. Schneider, Eds., vol. 12308.\hskip 1em
  plus 0.5em minus 0.4em\relax Springer, 2020, pp. 88--109.

\bibitem{roques2019detecting}
O.~Roques, S.~Maffeis, and M.~Cova, ``Detecting malware in tls traffic,'' in
  \emph{The IEEE Conference on Local Computer Networks 30th Anniversary
  (LCN’05)}, 2019.

\bibitem{DBLP:conf/iccns/DaiGLYLC19}
R.~Dai, C.~Gao, B.~Lang, L.~Yang, H.~Liu, and S.~Chen, ``{SSL} malicious
  traffic detection based on multi-view features,'' in \emph{{ICCNS} 2019: The
  9th International Conference on Communication and Network Security,
  Chongqing, China, November 15-17, 2019}.\hskip 1em plus 0.5em minus
  0.4em\relax {ACM}, 2019, pp. 40--46.

\bibitem{10.1007/978-3-030-67090-0_8}
W.~Lee and S.~Jin, ``Encrypted malware traffic detection using tls features and
  random forest,'' in \emph{Computational and Experimental Simulations in
  Engineering}, S.~N. Atluri and I.~Vu{\v{s}}anovi{\'{c}}, Eds.\hskip 1em plus
  0.5em minus 0.4em\relax Cham: Springer International Publishing, 2021, pp.
  85--100.

\bibitem{DBLP:conf/ccs/TorroledoCB18}
I.~Torroledo, L.~D. Camacho, and A.~C. Bahnsen, ``Hunting malicious {TLS}
  certificates with deep neural networks,'' in \emph{Proceedings of the 11th
  {ACM} Workshop on Artificial Intelligence and Security, {CCS} 2018, Toronto,
  ON, Canada, October 19, 2018}, S.~Afroz, B.~Biggio, Y.~Elovici, D.~Freeman,
  and A.~Shabtai, Eds.\hskip 1em plus 0.5em minus 0.4em\relax {ACM}, 2018, pp.
  64--73.

\bibitem{DBLP:conf/dsc/ZhaoLWHTC21}
C.~Zhao, S.~Li, X.~Wu, W.~Han, Z.~Tian, and M.~Chen, ``A novel malware
  encrypted traffic detection framework based on ensemble learning,'' in
  \emph{Sixth {IEEE} International Conference on Data Science in Cyberspace,
  {DSC} 2021, Shenzhen, China, October 9-11, 2021}.\hskip 1em plus 0.5em minus
  0.4em\relax {IEEE}, 2021, pp. 614--620.

\bibitem{DBLP:journals/virology/AndersonPM18}
B.~Anderson, S.~Paul, and D.~A. McGrew, ``Deciphering malware's use of {TLS}
  (without decryption),'' \emph{J. Comput. Virol. Hacking Tech.}, vol.~14,
  no.~3, pp. 195--211, 2018.

\bibitem{DBLP:conf/ccs/AndersonM16}
B.~Anderson and D.~A. McGrew, ``Identifying encrypted malware traffic with
  contextual flow data,'' in \emph{Proceedings of the 2016 {ACM} Workshop on
  Artificial Intelligence and Security, AISec@CCS 2016, Vienna, Austria,
  October 28, 2016}, D.~M. Freeman, A.~Mitrokotsa, and A.~Sinha, Eds.\hskip 1em
  plus 0.5em minus 0.4em\relax {ACM}, 2016, pp. 35--46.

\bibitem{gopal2018mitigating}
T.~S. Gopal, M.~Meerolla, G.~Jyostna, P.~R.~L. Eswari, and E.~Magesh,
  ``Mitigating mirai malware spreading in iot environment,'' in \emph{2018
  International Conference on Advances in Computing, Communications and
  Informatics (ICACCI)}.\hskip 1em plus 0.5em minus 0.4em\relax IEEE, 2018, pp.
  2226--2230.

\bibitem{garcia2015modelling}
S.~Garcia, ``Modelling the network behaviour of malware to block malicious
  patterns. the stratosphere project: a behavioural ips,'' \emph{Virus
  Bulletin}, pp. 1--8, 2015.

\bibitem{arp2015torben}
D.~Arp, F.~Yamaguchi, and K.~Rieck, ``Torben: A practical side-channel attack
  for deanonymizing tor communication,'' in \emph{Proceedings of the 10th ACM
  Symposium on Information, Computer and Communications Security}, 2015, pp.
  597--602.

\bibitem{DBLP:journals/ijics/GuptaS20}
A.~Gupta and L.~S. Sharma, ``A categorical survey of state-of-the-art intrusion
  detection system-\emph{Snort},'' \emph{Int. J. Inf. Comput. Secur.}, vol.~13,
  no. 3/4, pp. 337--356, 2020.

\bibitem{DBLP:conf/sca2/ChibaAMOR19}
Z.~Chiba, N.~Abghour, K.~Moussaid, A.~E. Omri, and M.~Rida, ``Newest
  collaborative and hybrid network intrusion detection framework based on
  suricata and isolation forest algorithm,'' in \emph{Proceedings of the 4th
  International Conference on Smart City Applications, {SCA} 2019, Casablanca,
  Morocco, October 02-04, 2019}.\hskip 1em plus 0.5em minus 0.4em\relax {ACM},
  2019, pp. 77:1--77:11.

\bibitem{DBLP:journals/tifs/DongLCLC21}
C.~Dong, Z.~Lu, Z.~Cui, B.~Liu, and K.~Chen, ``Mbtree: Detecting encryption
  rats communication using malicious behavior tree,'' \emph{{IEEE} Trans. Inf.
  Forensics Secur.}, vol.~16, pp. 3589--3603, 2021.

\bibitem{DBLP:journals/asc/MasdariK20}
M.~Masdari and H.~Khezri, ``A survey and taxonomy of the fuzzy signature-based
  intrusion detection systems,'' \emph{Appl. Soft Comput.}, vol.~92, p. 106301,
  2020.

\bibitem{DBLP:journals/cybersec/KhraisatGVK19}
A.~Khraisat, I.~Gondal, P.~Vamplew, and J.~Kamruzzaman, ``Survey of intrusion
  detection systems: techniques, datasets and challenges,'' \emph{Cybersecur.},
  vol.~2, no.~1, p.~20.

\bibitem{DBLP:conf/icct/IsingizweWLWWL21}
D.~F. Isingizwe, M.~Wang, W.~Liu, D.~Wang, T.~Wu, and J.~Li, ``Analyzing
  learning-based encrypted malware traffic classification with automl,'' in
  \emph{21st International Conference on Communication Technology, {ICCT} 2021,
  Tianjin, China, October 13-16, 2021}.\hskip 1em plus 0.5em minus 0.4em\relax
  {IEEE}, 2021, pp. 313--322.

\bibitem{10.1007/978-981-32-9949-8_48}
N.~A. Azeez, T.~M. Bada, S.~Misra, A.~Adewumi, C.~Van~der Vyver, and R.~Ahuja,
  ``Intrusion detection and prevention systems: An updated review,'' in
  \emph{Data Management, Analytics and Innovation}, N.~Sharma, A.~Chakrabarti,
  and V.~E. Balas, Eds.\hskip 1em plus 0.5em minus 0.4em\relax Singapore:
  Springer Singapore, 2020, pp. 685--696.

\bibitem{DBLP:conf/acling/ApplebaumG021}
S.~Applebaum, T.~Gaber, and A.~Ahmed, ``Signature-based and
  machine-learning-based web application firewalls: {A} short survey,'' in
  \emph{Fifth International Conference On Arabic Computational Linguistics,
  {ACLING} 2021, June 4-5, 2021, Virtual Event}, ser. Procedia Computer
  Science, K.~Shaalan and S.~R. El{-}Beltagy, Eds., vol. 189.\hskip 1em plus
  0.5em minus 0.4em\relax Elsevier, 2021, pp. 359--367.

\bibitem{DBLP:journals/jnsm/OtoumN21}
Y.~Otoum and A.~Nayak, ``{AS-IDS:} anomaly and signature based {IDS} for the
  internet of things,'' \emph{J. Netw. Syst. Manag.}, vol.~29, no.~3, p.~23,
  2021.

\bibitem{10.1007/978-3-030-52856-0_2}
N.~F. Firoz, M.~T. Arefin, and M.~R. Uddin, ``Performance optimization of
  layered signature based intrusion detection system using snort,'' in
  \emph{Cyber Security and Computer Science}, T.~Bhuiyan, M.~M. Rahman, and
  M.~A. Ali, Eds.\hskip 1em plus 0.5em minus 0.4em\relax Cham: Springer
  International Publishing, 2020, pp. 14--27.

\bibitem{bader2022maldist}
O.~Bader, A.~Lichy, C.~Hajaj, R.~Dubin, and A.~Dvir, ``Maldist: From encrypted
  traffic classification to malware traffic detection and classification,'' in
  \emph{2022 IEEE 19th Annual Consumer Communications \& Networking Conference
  (CCNC)}.\hskip 1em plus 0.5em minus 0.4em\relax IEEE, 2022, pp. 527--533.

\bibitem{niu2022novel}
Z.~Niu, J.~Xue, D.~Qu, Y.~Wang, J.~Zheng, and H.~Zhu, ``A novel approach based
  on adaptive online analysis of encrypted traffic for identifying malware in
  iiot,'' \emph{Information Sciences}, vol. 601, pp. 162--174, 2022.

\bibitem{fang2021communication}
Y.~Fang, K.~Li, R.~Zheng, S.~Liao, and Y.~Wang, ``A communication-channel-based
  method for detecting deeply camouflaged malicious traffic,'' \emph{Computer
  Networks}, vol. 197, p. 108297, 2021.

\bibitem{DBLP:journals/jsac/HanWZCYLSY21}
D.~Han, Z.~Wang, Y.~Zhong, W.~Chen, J.~Yang, S.~Lu, X.~Shi, and X.~Yin,
  ``Evaluating and improving adversarial robustness of machine learning-based
  network intrusion detectors,'' \emph{{IEEE} J. Sel. Areas Commun.}, vol.~39,
  no.~8, pp. 2632--2647, 2021.

\bibitem{DBLP:conf/ccs/AndresiniPPLAC21}
G.~Andresini, F.~Pendlebury, F.~Pierazzi, C.~Loglisci, A.~Appice, and
  L.~Cavallaro, ``{INSOMNIA:} towards concept-drift robustness in network
  intrusion detection,'' in \emph{AISec@CCS 2021: Proceedings of the 14th {ACM}
  Workshop on Artificial Intelligence and Security, Virtual Event, Republic of
  Korea, 15 November 2021}, N.~Carlini, A.~Demontis, and Y.~Chen, Eds.\hskip
  1em plus 0.5em minus 0.4em\relax {ACM}, 2021, pp. 111--122.

\bibitem{DBLP:conf/ndss/EdeBCRDLCSP20}
T.~van Ede, R.~Bortolameotti, A.~Continella, J.~Ren, D.~J. Dubois,
  M.~Lindorfer, D.~R. Choffnes, M.~van Steen, and A.~Peter, ``Flowprint:
  Semi-supervised mobile-app fingerprinting on encrypted network traffic,'' in
  \emph{27th Annual Network and Distributed System Security Symposium, {NDSS}
  2020, San Diego, California, USA, February 23-26, 2020}.\hskip 1em plus 0.5em
  minus 0.4em\relax The Internet Society, 2020.

\bibitem{DBLP:conf/eurosp/KrawczykW16}
H.~Krawczyk and H.~Wee, ``The {OPTLS} protocol and {TLS} 1.3,'' in \emph{{IEEE}
  European Symposium on Security and Privacy, EuroS{\&}P 2016,
  Saarbr{\"{u}}cken, Germany, March 21-24, 2016}.\hskip 1em plus 0.5em minus
  0.4em\relax {IEEE}, 2016, pp. 81--96.

\bibitem{DBLP:journals/network/DainottiPC12}
A.~Dainotti, A.~Pescap{\`{e}}, and K.~C. Claffy, ``Issues and future directions
  in traffic classification,'' \emph{{IEEE} Netw.}, vol.~26, no.~1, pp. 35--40,
  2012.

\bibitem{DBLP:journals/tifs/ShenZZXD21}
M.~Shen, J.~Zhang, L.~Zhu, K.~Xu, and X.~Du, ``Accurate decentralized
  application identification via encrypted traffic analysis using graph neural
  networks,'' \emph{{IEEE} Trans. Inf. Forensics Secur.}, vol.~16, pp.
  2367--2380, 2021.

\bibitem{DBLP:conf/nips/MnihHGK14}
V.~Mnih, N.~Heess, A.~Graves, and K.~Kavukcuoglu, ``Recurrent models of visual
  attention,'' in \emph{Advances in Neural Information Processing Systems 27:
  Annual Conference on Neural Information Processing Systems 2014, December
  8-13 2014, Montreal, Quebec, Canada}, Z.~Ghahramani, M.~Welling, C.~Cortes,
  N.~D. Lawrence, and K.~Q. Weinberger, Eds., 2014, pp. 2204--2212.

\bibitem{DBLP:conf/nips/VaswaniSPUJGKP17}
A.~Vaswani, N.~Shazeer, N.~Parmar, J.~Uszkoreit, L.~Jones, A.~N. Gomez,
  L.~Kaiser, and I.~Polosukhin, ``Attention is all you need,'' in
  \emph{Advances in Neural Information Processing Systems 30: Annual Conference
  on Neural Information Processing Systems 2017, December 4-9, 2017, Long
  Beach, CA, {USA}}, I.~Guyon, U.~von Luxburg, S.~Bengio, H.~M. Wallach,
  R.~Fergus, S.~V.~N. Vishwanathan, and R.~Garnett, Eds., 2017, pp. 5998--6008.

\bibitem{DBLP:conf/kes/GittinsS20}
Z.~Gittins and M.~Soltys, ``Malware persistence mechanisms,'' in
  \emph{Knowledge-Based and Intelligent Information {\&} Engineering Systems:
  Proceedings of the 24th International Conference KES-2020, Virtual Event,
  16-18 September 2020}, ser. Procedia Computer Science, M.~Cristani, C.~Toro,
  C.~Zanni{-}Merk, R.~J. Howlett, and L.~C. Jain, Eds., vol. 176.\hskip 1em
  plus 0.5em minus 0.4em\relax Elsevier, 2020, pp. 88--97.

\bibitem{kuraku2020emotet}
S.~Kuraku and D.~Kalla, ``Emotet malware—a banking credentials stealer,''
  \emph{Iosr J. Comput. Eng}, vol.~22, pp. 31--41, 2020.

\bibitem{DBLP:conf/icadiwt/KhanRAFSV14}
K.~Khan, S.~ur~Rehman, K.~Aziz, S.~Fong, S.~Sarasvady, and A.~Vishwa,
  ``{DBSCAN:} past, present and future,'' in \emph{The Fifth International
  Conference on the Applications of Digital Information and Web Technologies,
  {ICADIWT} 2014, Chennai, India, February 17-19, 2014}.\hskip 1em plus 0.5em
  minus 0.4em\relax {IEEE}, 2014, pp. 232--238.

\bibitem{DBLP:journals/corr/abs-1901-09069}
F.~Almeida and G.~Xex{\'{e}}o, ``Word embeddings: {A} survey,'' \emph{CoRR},
  vol. abs/1901.09069, 2019.

\bibitem{DBLP:journals/corr/abs-1301-3781}
T.~Mikolov, K.~Chen, G.~Corrado, and J.~Dean, ``Efficient estimation of word
  representations in vector space,'' in \emph{1st International Conference on
  Learning Representations, {ICLR} 2013, Scottsdale, Arizona, USA, May 2-4,
  2013, Workshop Track Proceedings}, Y.~Bengio and Y.~LeCun, Eds., 2013.

\bibitem{stratodatasets}
Stratosphere, ``Stratosphere laboratory datasets,'' 2015, retrieved March 13,
  2020, from \url{https://www.stratosphereips.org/datasets-overview}.

\bibitem{DBLP:conf/icissp/LashkariDMG17}
A.~H. Lashkari, G.~Draper{-}Gil, M.~S.~I. Mamun, and A.~A. Ghorbani,
  ``Characterization of tor traffic using time based features,'' in
  \emph{Proceedings of the 3rd International Conference on Information Systems
  Security and Privacy, {ICISSP} 2017, Porto, Portugal, February 19-21, 2017},
  P.~Mori, S.~Furnell, and O.~Camp, Eds.\hskip 1em plus 0.5em minus 0.4em\relax
  SciTePress, 2017, pp. 253--262.

\bibitem{DBLP:conf/icissp/Draper-GilLMG16}
G.~Draper{-}Gil, A.~H. Lashkari, M.~S.~I. Mamun, and A.~A. Ghorbani,
  ``Characterization of encrypted and {VPN} traffic using time-related
  features,'' in \emph{Proceedings of the 2nd International Conference on
  Information Systems Security and Privacy, {ICISSP} 2016, Rome, Italy,
  February 19-21, 2016}, O.~Camp, S.~Furnell, and P.~Mori, Eds.\hskip 1em plus
  0.5em minus 0.4em\relax SciTePress, 2016, pp. 407--414.

\bibitem{lashkari2017cicflowmeter}
A.~H. Lashkari, Y.~Zang, G.~Owhuo, M.~Mamun, and G.~Gil, ``Cicflowmeter,''
  2017.

\bibitem{271335}
K.~Cho, K.~Mitsuya, and A.~Kato, ``Traffic data repository at the {WIDE}
  project,'' in \emph{2000 USENIX Annual Technical Conference (USENIX ATC
  00)}.\hskip 1em plus 0.5em minus 0.4em\relax San Diego, CA: USENIX
  Association, Jun. 2000.

\bibitem{malware-traffic-analysis}
\BIBentryALTinterwordspacing
B.~Duncan, ``Malware traffic analysis.'' [Online]. Available:
  \url{http://malware-traffic-analysis.net}
\BIBentrySTDinterwordspacing

\end{thebibliography}

\end{document}